\documentclass[11pt,a4paper]{article}
\usepackage[utf8]{inputenc}
\usepackage{amsmath}
\usepackage{amsfonts}
\usepackage{amssymb}
\usepackage[left=2cm,right=2cm,top=2cm,bottom=2cm]{geometry}

\usepackage{slashed}
\usepackage{hyperref}
\usepackage{graphicx}
\usepackage{caption}
\usepackage{float}
\usepackage{subcaption}
\usepackage{multirow}
\usepackage[dvipsnames]{xcolor}
\usepackage{multicol}
\usepackage{authblk}

\renewcommand{\theequation}{\arabic{section}.\arabic{equation}}

\usepackage{ifthen}
\usepackage{tikz}
\usepackage{xspace}
\usepackage{listings}
\usepackage[T1]{fontenc}

\definecolor{mediumjunglegreen}{rgb}{0.11, 0.21, 0.18}
\definecolor{bg}{HTML}{282828}

\newcommand*\cppin{\lstinline[language=c++]}

\newcommand*\pyin{\lstinline[language=c++]}

\lstnewenvironment{cpp}
{\lstset{language=c++}}
{}

\lstnewenvironment{py}
{\lstset{language=Python}}
{}

\lstnewenvironment{bash}
{\lstset{language=Python}}
{}

\lstnewenvironment{pseudo}
{\lstset{language=Python}}
{}

\newenvironment{run}[1]{``{\tt #1}"\xspace}

\newcommand{\dint}{  \displaystyle \int }
\newcommand{\ie}{{\em i.e.}\xspace}
\newcommand{\eg}{{\em e.g.}\xspace}
\newcommand{\GeV}{{\rm GeV}\xspace}

\newcommand{\MeV}{{\rm MeV}\xspace}

\def\mimes{{\tt MiMeS}\xspace}

\newcommand{\CPP}{{\tt C++}\xspace}

\newcommand{\PY}{{\tt python}\xspace}

\newcommand{\JUPY}{{\tt jupyter}\xspace}

\newcommand{\geff}{ g_{\rm eff}{}\xspace}
\newcommand{\heff}{ h_{\rm eff}{}\xspace}

\newcommand{\Ham}{ \mathcal{H}\xspace}

\newcommand{\thetamax}{ \theta_{\rm peak}{}\xspace}

\newcommand{\thetai}{ \theta_{\rm ini}{}\xspace}

\newcommand{\fa}{ f_{a}{}\xspace}

\newcommand{\ti}{ t_{\rm ini}{}\xspace}

\newcommand{\ai}{ a_{\rm ini}{}\xspace}

\newcommand{\thetaosc}{ \theta_{\rm osc}{}\xspace}

\newcommand{\Tosc}{ T_{\rm osc}{}\xspace}

\newcommand{\tosc}{ t_{\rm osc}{}\xspace}

\newcommand{\aosc}{ a_{\rm osc}{}\xspace}

\newcommand{\Omegai}{ \Omega_{\rm ini}\xspace}

\newcommand{\ma}{ m_{a}{}\xspace}
\newcommand{\maT}{ \tilde m_{a}{}\xspace}

\newcommand{\lrb}[1]{\left( #1 \right)}
\newcommand{\lrsb}[1]{\left[ #1 \right]}
\newcommand{\lrBigb}[1]{\Big( #1 \Big)}

\newcommand{\lrBiggb}[1]{\Bigg( #1 \Bigg)}

\newcounter{NumArgs}

\newcommand{\eqs}[1]{\setcounter{NumArgs}{0}\foreach\i in{#1}{\stepcounter{NumArgs}}%
	\ifthenelse{\equal{\theNumArgs}{1}}{eq.~(\ref{#1})}%
	{\ifthenelse{\equal{\theNumArgs}{2}}%
		{eqs.~\foreach\i[count=\q]in{#1}{\ifthenelse{\equal{\q}{\theNumArgs}}{and (\ref{\i})}{(\ref{\i})~}}}%
		{eqs.~\foreach\i[count=\q]in{#1}{\ifthenelse{\equal{\q}{\theNumArgs}}{and (\ref{\i})}{(\ref{\i}),~}}}}}

\newcommand{\Eqs}[1]{\setcounter{NumArgs}{0}\foreach\i in{#1}{\stepcounter{NumArgs}}%
	\ifthenelse{\equal{\theNumArgs}{1}}{Eq.~(\ref{#1})}%
	{\ifthenelse{\equal{\theNumArgs}{2}}%
		{Eqs.~\foreach\i[count=\q]in{#1}{\ifthenelse{\equal{\q}{\theNumArgs}}{and (\ref{\i})}{(\ref{\i})~}}}%
		{Eqs.~\foreach\i[count=\q]in{#1}{\ifthenelse{\equal{\q}{\theNumArgs}}{and (\ref{\i})}{(\ref{\i}),~}}}}}

\newcommand{\refs}[1]{\setcounter{NumArgs}{0}\foreach\i in{#1}{\stepcounter{NumArgs}}%
	\ifthenelse{\equal{\theNumArgs}{1}}{(\ref{#1})}%
	{\ifthenelse{\equal{\theNumArgs}{2}}%
		{\foreach\i[count=\q]in{#1}{\ifthenelse{\equal{\q}{\theNumArgs}}{and (\ref{\i})}{(\ref{\i})~}}}%
		{\foreach\i[count=\q]in{#1}{\ifthenelse{\equal{\q}{\theNumArgs}}{and (\ref{\i})}{(\ref{\i}),~}}}}}

\newcommand{\Figs}[1]{\setcounter{NumArgs}{0}\foreach\i in{#1}{\stepcounter{NumArgs}}%
	\ifthenelse{\equal{\theNumArgs}{1}}{Fig.~(\ref{#1})}%
	{\ifthenelse{\equal{\theNumArgs}{2}}%
		{Figs.~\foreach\i[count=\q]in{#1}{\ifthenelse{\equal{\q}{\theNumArgs}}{and (\ref{\i})}{(\ref{\i})~}}}%
		{Figs.~\foreach\i[count=\q]in{#1}{\ifthenelse{\equal{\q}{\theNumArgs}}{and (\ref{\i})}{(\ref{\i}),~}}}}}

\newcommand{\Gen}[2]{\setcounter{NumArgs}{0}\foreach\i in{#2}{\stepcounter{NumArgs}}%
	\ifthenelse{\equal{\theNumArgs}{1}}{#1.~(\ref{#2})}%
	{\ifthenelse{\equal{\theNumArgs}{2}}%
		{#1.~\foreach\i[count=\q]in{#2}{\ifthenelse{\equal{\q}{\theNumArgs}}{and (\ref{\i})}{(\ref{\i})~}}}%
		{#1.~\foreach\i[count=\q]in{#2}{\ifthenelse{\equal{\q}{\theNumArgs}}{and (\ref{\i})}{(\ref{\i}),~}}}}}

\author[ ]{Karamitros Dimitrios}
\affil[ ]{\em School of Physics and Astronomy, The University of Manchester,
	Manchester M13 9PL, United Kingdom}
\affil[ ]{\textit{E-mail: } \href{mailto:dimitrios.karamitros@manchester.ac.uk}{\color{blue}{dimitrios.karamitros@manchester.ac.uk}}}

\title{{\tt MiMeS}: Misalignment Mechanism Solver}
\begin{document}

\maketitle

\begin{abstract}
	We introduce a \CPP header-only library that is used to solve the axion equation of motion, \mimes.  
	\mimes makes no assumptions regarding the cosmology and the mass of the axion, which allows the user 
	to consider various cosmological scenarios and axion-like models.
	\mimes also includes a convenient \PY interface that allows the library to be called without writing any code in \CPP, with minimal overhead.
\end{abstract}

{\bf Program summary:}

{\sl 
	Program title: \mimes.
	
	Developer's respository link: \href{https://github.com/dkaramit/MiMeS}{https://github.com/dkaramit/MiMeS}.
	
	Programming language: \CPP and \PY.
	
	Licensing provisions: MIT license.
	
	Nature of problem: Solving numerically the axion (or axion-like-particle) equation of motion, in order to determine the corresponding relic abundance.  The library is designed to be quite general, 
	and can be used to obtain the relic abundance in various cosmological scenarios, and axion-like-particle models. 
	
	Solution method: Embedded Runge-Kutta for the numerical integration of the equation of motion. The user may choose between explicit and Rosenborck methods, or implement their own Butcher tableau. For the various interpolations, the library uses cubic splines. 
	
	Restrictions: The the derivative of the axion-angle initially is assumed vanish. This is hard-coded in the library, and there is no easy way for the user to change it. Furthermore, any additional contribution from decays or annihilation of plasma particles to the axion (or ALP) energy density is assumed to be subdominant.   
}

\tableofcontents

\section{Introduction}\label{sec:intro}
\setcounter{equation}{0}
The axion is a hypothetical particle that was originally introduced in order to solve the strong CP-problem of the standard model (SM)~\cite{Peccei:1977hh,Weinberg:1977ma,Wilczek:1977pj}. Furthermore, the axion is assumed to acquire a non-zero vacuum expectation value (VEV), which results in the spontaneous breaking of a global symmetry, called Peccei-Quinn (PQ). That is the axion is a pseudo-Nambu-Goldstone boson.  Moreover, it also appears to be a valid dark matter (DM) candidate~\cite{Preskill:1982cy,Dine:1982ah,Abbott:1982af}, as it is electrically neutral and long-lived.  The axion starts, at very early time, with a VEV close to the PQ breaking scale (usually much higher than $100\GeV$), but due to the expansion of the Universe it eventually ends oscillating around zero. This oscillation results in an apparent constant number of axion particles (or constant energy of the axion field) today, which can account for the observed~\cite{Planck:2018vyg} DM relic abundance. 
Apart form the axion, there are other hypothetical particles (for early examples of such particles see~\cite{Chikashige:1980ui,Georgi:1981pu}; and~\cite{Ringwald:2014vqa} for a review), called axion-like particles (ALPs), which are not related to the strong CP-problem. That is, these ALPs, interact with the SM in a different way than the axion. However, they can still account for the DM of the Universe, and both axions and ALPs, follow the similar dynamics during the early universe; the follow the same equation of motion (EOM), with a different mass. The mass of the axion is dictated by QCD, while it is different (model specific) for ALPs.  Therefore, the cosmological evolution of both axions and ALPs is often discussed together (see, for example ref.~\cite{MARSH20161}). 

Moreover, deviation from the standard cosmological evolution is possible, as long as any non-standard contribution to the energy density is absent before Big Bang Nucleosynthesis~\cite{Kolb:206230,Peebles:1993} becomes active; for temperatures~\cite{Kawasaki:2000en, Hannestad:2004px, Ichikawa:2005vw, DeBernardis:2008zz} $T\gtrsim \mathcal{O}(10)~\MeV$. For the QCD axion, such studies have been performed~\cite{Visinelli:2009kt, Arias:2021rer}. However, updated experimental data or new ALP models may require more similar systematic studies. To the best of our knowledge, a library for the calculation of the axion (or ALP) relic abundance does not exist. 

Therefore, in this article, we introduce \mimes;  a header-only \CPP library,~\footnote{\mimes is distributed under the MIT license. A copy of this license should be available in the \mimes root directory. If you have not received one, you can find it at  \href{https://github.com/dkaramit/MiMeS/blob/master/LICENSE}{github.com/dkaramit/MiMeS/blob/master/LICENSE}.} that solves the axion (or ALP) EOM, where both the mass and the underlying cosmological evolution are treated as user inputs. That is, \mimes can be used to compute the relic abundance of axions or ALPs, in a wide variety of scenarios.

There are several advantages of having a library that can calculate the relic abundance -- in principle -- fast. For example, one can perform a scan over different cosmological scenarios for various ALPs cases, automatically, writing only a few lines of code. Furthermore, the availability of a tool can help the community reproduce published results, without spending time duplicating the overall effort. \mimes aims to be a useful addition in the list of available computational tools for physics,~\footnote{Especially for dark matter, where tools for both thermal and non-thermal production have been developed~\cite{Gondolo:2004sc,Belanger:2013oya,Belanger:2018ccd,Binder:2021bmg}; however, without the addition of the misalignment mechanism.} since it is designed to be simple to use, simple to understand, and simple to modify.

This article is organized as follows: 
In section~\ref{sec:Physics}, we introduce the EOM and show how it can be solved approximately. Also, we introduce the notation that \mimes follows, we derive the ``adiabatic invariant" of the system, and discuss how \mimes uses it. 
In the next section, we introduce \mimes,
 by showing how it can be downloaded, compiled, and run for the first time. Also, we explain in detail all the parameters that \mimes as a user input at run-time.
In section~\ref{sec:assumptions}, we discuss the few assumption that \mimes makes, its default compile-time options and user input, and how the user can change them. Moreover, we provide a complete example in both \CPP and \PY. 

\section{Physics background}\label{sec:Physics}
\setcounter{equation}{0}
Although there are several works in the literature (such as~\cite{Chang:1998ys,MARSH20161}) that can provide an insight on the cosmological evolution of axions, in this section we define, derive, and discuss various quantities we need, in order to understand how \mimes works in detail.

\paragraph{The EOM} 
The axion field, $A$, is usually expressed in terms of the so-called axion angle, $\theta$, as $A = \theta \ \fa$, with $\fa$ the scale at which the PQ symmetry breaks.~\footnote{If in a model under study, there is no $\fa$, this parameter is still expected by \mimes, but the user can set $\fa=1$.} 
The axion angle follows the EOM 
\begin{equation}
	\lrb{\dfrac{d^2}{d t^2} + 3 H(t) \ \dfrac{d}{d t} } \theta(t) + \maT^2(t) \ \sin \theta(t) = 0 \; ,
	\label{eq:eom}
\end{equation}
with $H(t)$ the Hubble parameter (determined by the cosmology), and $\maT(t)$  the time (temperature) dependent mass of the axion. Usually the axion mass is written as
\begin{equation}
	\maT^2(T) = \dfrac{\chi(T)}{\fa^2}\;,
	\label{eq:axion_mass_def}
\end{equation} 
with $\chi$ a function of the temperature. For the QCD axion, this has been calculated using lattice simulations in~\cite{Borsanyi:2016ksw}. \mimes comes with the data provided by ref.~\cite{Borsanyi:2016ksw}. However, the user is free to change them, or use another function for the mass.

\paragraph{Initial conditions}
\begin{figure}[h!]
	\includegraphics[width=1\textwidth]{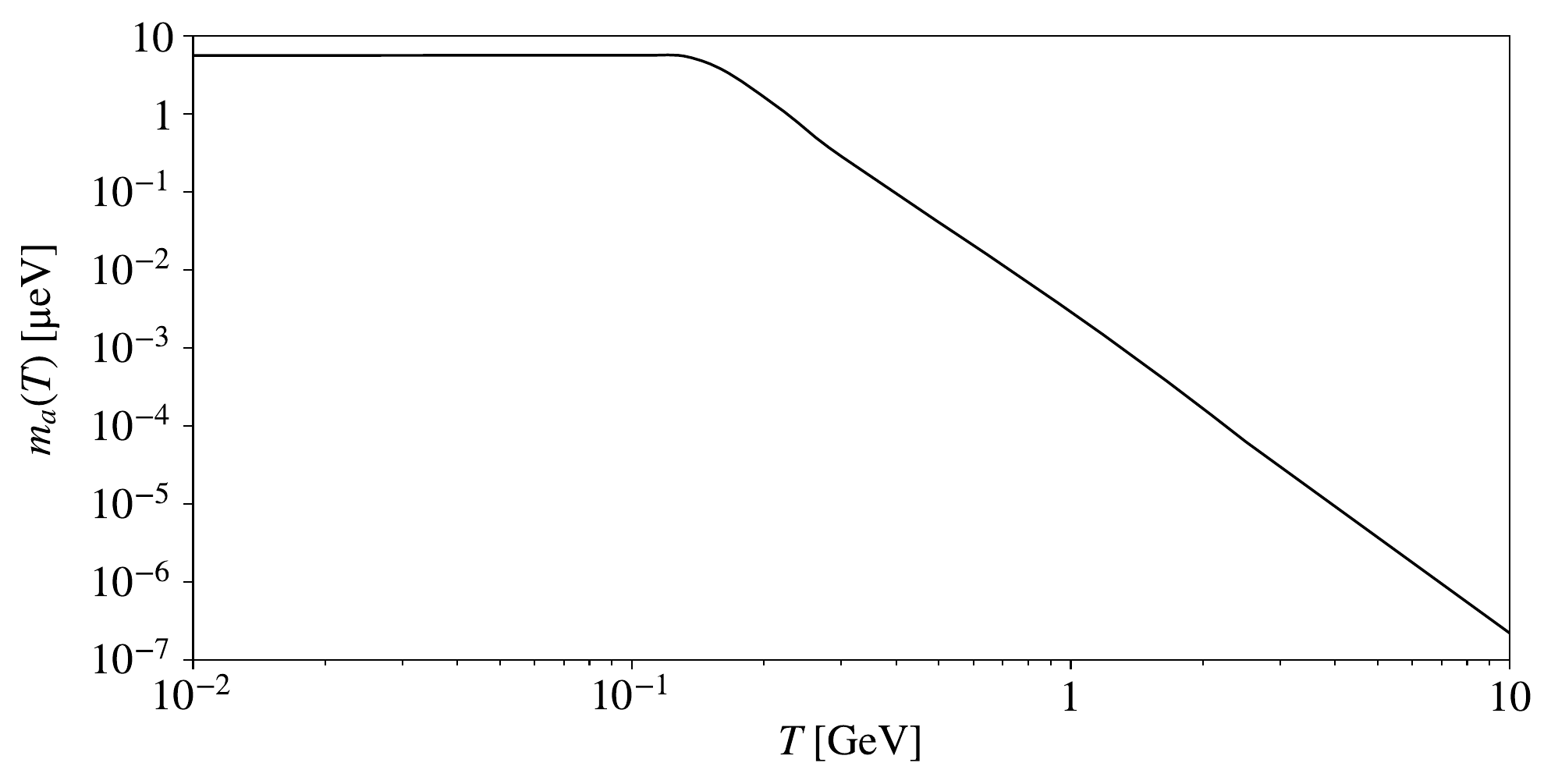}
	\caption{The mass of the axion as a function of the temperature for $\fa=10^{12}~\GeV$, using the data provided in ref.~\cite{Borsanyi:2016ksw}.}
	\label{fig:axion_mass}
\end{figure}
We assume that at very high temperatures $\maT \to 0$ (for the QCD axion this is true; see \Figs{fig:axion_mass}) -- \ie $\maT \ll H$ -- at very early times.  Therefore, at the very early Universe, the EOM is\footnote{For the QCD axion the PQ symmetry breaking scale determines whether there are domains with different $\thetai$.  However, one can still use a mean value of the angle as its initial condition. This is discussed in the literature; \eg in~\cite{Preskill:1982cy,Visinelli:2009zm,Visinelli:2009kt}.}
\begin{equation}
	\lrb{\dfrac{d^2}{d t^2} + 3 H(t) \ \dfrac{d}{d t}  } \theta(t) = 0 \; .
	\label{eq:massless_eom}
\end{equation}
which is solved by $\theta = \thetai + C \dint_{0}^t d t' \ \lrb{ \dfrac{a(t'=0)}{a(t')} }^3$, with $\thetai$ a constant (some initial value) and $a$ the scale factor of the Universe. That is, as the Universe expands, and as long as $\maT \ll H$, $\theta \to \thetai$, and $\dot \theta \to 0$. Since we would like to calculate the angle today, we can integrate \eqs{eq:eom} from a time after inflation (call it $t = \ti$) such that $ \dot \theta |_{t=\ti} = 0$ and  $\theta|_{t=\ti} = \thetai$.   This is the most common case (there are exceptions to this; \eg~\cite{Co:2019jts}), and it is what \mimes uses.

\paragraph{The shift symmetry}
It is important to note that  $\dot \theta \sim a^{-3}$ is expected from symmetry grounds. The axion is assumed to be the phase of a complex scalar field
\begin{equation}
	\Phi(x) = \dfrac{1}{\sqrt{2}}(\fa + \phi(x)) \exp{(i \theta(x))} \;, 
\end{equation}
charged under a global symmetry. If this symmetry is not broken, then the Lagrangian is invariant under a shift $\theta \to \theta + z$. Then there exists a Noether charge $Q  \sim a^3 \dot \theta = const.$ Thus, at high temperatures tha shift symmetry is restored (\ie the mass of the axion vanishes is much smaller than its kinetic energy), and $\dot \theta \sim a^{-3}$.

On the other hand, if at high temperatures there exist shift symmetry  breaking interactions, then $ a^3 \dot \theta \neq const.$ This means that at low temperatures, $\dot \theta$ can be sizeable. The explicit breaking of the shift symmetry was introduced recently as a means to alter the axion production mechanism  (for details on how this affects the evolution of the axion see \eg  ~\cite{Co:2019jts,Chang:2019tvx}). Practically, a non-vanishing $\dot \theta$ can change the point where it starts to oscillate, which greatly affects its final energy density. 

From this point, we are not going to discuss a non-vanishing $\dot \theta |_{t=\ti}$  further, since it is beyond the scope of \mimes.

\subsection{The WKB approximation}
In order to solve analytically \eqs{eq:eom}, we assume $\theta \ll 1$, which results in the linearised EOM
\begin{equation}
	\lrb{\dfrac{d^2}{d t^2} + 3 H(t) \ \dfrac{d}{d t} + \maT^2(t) } \theta(t) = 0 \; .
	\label{eq:linear_eom}
\end{equation}

Using a trial solution $\theta_{\rm trial} = \exp\lrsb{ i \dint d t \ \lrBigb{\psi(t) +3/2 \ i \ H(t)} }$, and defining $\Omega^2 = \maT^2 - \dfrac{9}{4} H^2 -  \dfrac{3}{2} \dot H $ we can transform the \eqs{eq:linear_eom} to 
\begin{equation}
	\psi^2 = \Omega^2 + i \ \dot \psi \; ,
	\label{eq:eom_of_psi}
\end{equation}
which has a formal solution $\psi = \pm \sqrt{\Omega^2 + i \dot \psi}$. In the WKB approximation, we assume a slow time-dependence; that is $\dot \psi \ll \Omega^2$ and $\dot \Omega \ll \Omega^2$. Then we can approximate $\psi$ as
\begin{equation}
	\psi \approx \pm \Omega + \dfrac{i}{2} \dfrac{d \log \Omega}{d t} \;,
	\label{eq:u_approx}
\end{equation}
which results in the general solution of \eqs{eq:linear_eom} 
\begin{equation}
	\theta \approx \dfrac{1}{\sqrt{\Omega}} \exp\lrb{-\dfrac{3}{2} \int d t \ H} \lrsb{ A \cos\lrb{ \int d t \ \Omega} +  B \sin\lrb{ \int d t \ \Omega}    } \;. 
	\label{eq:general_solution_eom_approx}
\end{equation}

Applying, then, the initial conditions $ \dot \theta |_{t=\ti} = 0$ and  $\theta|_{t=\ti} \approx \thetai$, we arrive at 
\begin{equation}
\theta(t) \approx \thetai \sqrt{ \dfrac{ \Omegai }{\Omega (t)} } \lrb{\dfrac{a}{\ai}}^{-3/2} \  \cos\lrb{ \int_{\ti}^t d t^\prime  \ \Omega(t^\prime)}   \;.
\label{eq:solution_eom_approx} 
\end{equation}

In order to further simplify this approximate result, we note that $\theta$ deviates from $\thetai$ close to $t=\tosc$ -- corresponding to $T = \Tosc$, the so-called ``oscillation temperature" -- $\maT|_{t = \tosc} = 3 H|_{t = \tosc}$, which is defined as the point at which the axion begins to oscillate. 
This observation allows us to set $\ti = \tosc$.  Moreover, at $t > \tosc$, we approximate $\Omega \approx \maT$, as $H^2$ and $\dot H$ become much smaller than $\maT^2$ quickly after $t=\tosc$. Finally, the axion angle takes the form
\begin{equation}
	\theta(t) \approx \thetaosc \lrb{\dfrac{3}{4}}^{1/4} \sqrt{ \dfrac{ \maT|_{t=\tosc} }{\maT  (t)} } \lrb{\dfrac{a}{\aosc}}^{-3/2} \  \cos\lrb{ \int_{\tosc}^t d t^\prime  \ \maT(t^\prime)}   \;,
	\label{eq:solution_eom_approx_theta_osc} 
\end{equation}
where $\thetaosc = \theta|_{t=\tosc}$. This equation is further simplified if we assume that $\thetaosc \approx \thetai$, \ie
\begin{equation}
	\theta(t) \approx \thetai \lrb{\dfrac{3}{4}}^{1/4} \sqrt{ \dfrac{ \maT|_{t=\tosc} }{\maT  (t)} } \lrb{\dfrac{a}{\aosc}}^{-3/2} \  \cos\lrb{ \int_{\tosc}^t d t^\prime  \ \maT(t^\prime)}   \;.
	\label{eq:solution_eom_approx_final} 
\end{equation}
It is worth mentioning that the accuracy of this approximation depends, in general, on $\Tosc$; it determines the difference between $\thetai$ and $\thetaosc$, the deviation of $\dot \theta|_{t=\tosc}$ from $0$, and whether $\dot \Omega \ll \Omega^2$.

\paragraph{Axion energy density}
In the small angle approximation, the energy density of the axion is 
\begin{eqnarray}
	\rho_{a} = \dfrac{1}{2} \fa^2 \lrsb{ \dot{\theta}^2 + \maT^2 \theta^2 } \;.
	\label{eq:rho_a_def} 
\end{eqnarray}
For the relic abundance of axions, we need to calculate their energy density at very late times. That is, $\dot{\tilde{m}}_a = 0$, $\maT \gg H$ and $\dot H \ll H^2$. After some algebra, we obtain the approximate form of the energy density as a function of the scale factor 
\begin{eqnarray}
	\rho_{a} \approx \dfrac{\ma }{2}  \ \fa^2 \ \thetai^2  \ \maT(\aosc) \ \lrb{\dfrac{\aosc}{a}}^3 \;,
	\label{eq:rho_a0} 
\end{eqnarray}
which shows that the energy density of axions at late times scales as the energy density of matter; \ie the number of axion particles is conserved. If there is a period of entropy injection to the plasma for $T<\Tosc$, the axion energy density gets diluted, since 
\begin{equation}
	a^3 \ s = \gamma \ \aosc^3 \ s_{\rm osc} \Rightarrow  \lrb{\dfrac{\aosc}{a}}^3 = \gamma^{-1} \dfrac{s}{s_{\rm osc}} \;,
	\label{eq:WKB_gamma_def}
\end{equation}
with $\gamma$ the amount of entropy injection to the plasma between $\tosc$ and $t$. Therefore, the present (at $T=T_0$) energy density of the axion, becomes
\begin{eqnarray}
	\rho_{a,0} = \gamma^{-1}  \dfrac{s_0}{s_{\rm osc}} \  \dfrac{1 }{2}  \ \fa^2 \ \ma \ \maT_{,{\rm osc}} \ \thetai^2    \;,
	\label{eq:rho_a_approx} 
\end{eqnarray}
with $\ma$ the mass of the axion at $T=T_0$. Notice that the explicit dependence on $\fa$ cancels if $\maT=\sqrt{\chi(T)}/\fa$. That is, $\fa$ only affects the energy density of the axions through its impact on $\Tosc$. 

\subsection{Notation}\label{sec:notation}
%
%
The EOM~(\ref{eq:eom}) depends on time, which is not useful variable in cosmology, especially non-standard cosmologies. Therefore, we introduce 
\begin{eqnarray}
	u = \log \dfrac{a}{\ai} \;,
	\label{eq:natation}
\end{eqnarray}
which results in 
\begin{eqnarray}
	&\dfrac{d F}{dt} &=  H  \dfrac{d F}{du} 
	\nonumber \\
	&\dfrac{d^2 F}{dt^2} &= H^2 \ \lrb{ \dfrac{d^2 F}{du^2} + \dfrac{1}{2} \dfrac{d \log H^2}{du}  \dfrac{d F}{du} }\;.
	\label{eq:deriv_u}
\end{eqnarray}
The EOM in terms of $u$, then, becomes
\begin{equation}
	\dfrac{d^2  \theta}{du^2} + \lrsb{\dfrac{1}{2} \dfrac{d \log H^2}{du} + 3 } \dfrac{d  \theta}{d u} + \ \lrb{\dfrac{\maT}{H}}^2 \ \sin \theta
	=0 \;.
	\label{eq:eom_u}
\end{equation}
Notice that in a radiation dominated Universe
$$
\dfrac{d \log H^2}{du} = -\lrb{ \dfrac{d \log \geff}{d \log T} +4 } \delta_h^{-1}\;,
$$
with  $ \delta_h = 1+ \dfrac{1}{3} \dfrac{d \log \heff}{d \log T} $. 
In a general cosmological setting, if the expansion rate is dominated by an energy density that scales as $\rho \sim a^{-c}$, $\dfrac{d \log H^2}{du}  = -c$. 
We notice that close to rapid particle annihilations and decays, the evolution of the energy densities change, and $\dfrac{d \log H^2}{du}$ can only be computed numerically.

Moreover, it is worth mentioning that the EOM~\ref{eq:eom_u} with the initial condition $\theta(u=0)=\thetai$ and $d\theta/du (u=0)=0$ can be written as a system of first 
order ordinary differential equations
\begin{eqnarray}
& \dfrac{d  \zeta}{du} + \lrsb{\dfrac{1}{2} \dfrac{d \log H^2}{du} + 3 } \zeta + \ \lrb{\dfrac{\maT}{H}}^2 \ \sin \theta
=0 \;.  \nonumber \\
&\dfrac{d \theta}{d u} - \zeta=0 \;.
\label{eq:eom_sys}
\end{eqnarray}
This form of the EOM is what \mimes uses, as it is suitable for integration via Runge-Kutta (RK) methods -- which are briefly discussed in Appendix~\ref{app:RK}.

\subsection{Adiabatic invariant and the anharmonic factor}\label{sec:an_fac}
The EOM~\ref{eq:eom} ca be solved analytically in the approximation $\theta \ll 1$. Moreover, even for $\theta\ll 1$, the WKB approximation fails to capture the dynamics before the adiabatic conditions are met, and result in an inaccurate axion relic abundance.  Therefore, a numerical integration should be preferred. Furthermore, in order to reduce the computation time, the numerical integration needs to stop as soon as the axion begins to evolve adiabatically. After this point, we can correlate its energy density at later times, using an ``adiabatic invariant", which can be defined  for oscillatory systems with varying period.

\paragraph{Definition of the adiabatic invariant}
Given a system with Hamiltonian $\mathcal{H}(\theta,p;t)$, the equations of motion are 
\begin{equation}
	\dot p = - \dfrac{\partial \mathcal{H}}{\partial \theta} \;, \;\; 
	\dot \theta =  \dfrac{\partial \mathcal{H}}{\partial p} \;.
	\label{eq:hamiltonian_eoms}
\end{equation}
Moreover, we note that
\begin{equation}
	d \Ham = \dot \theta \ d p - \dot p \ d \theta + \dfrac{\partial \Ham}{\partial t} \ d t \;.  
	\label{eq:total_dH}
\end{equation}
If this system exhibits closed orbits (\eg if it oscillates), we define 
\begin{equation}
	J \equiv C \ \oint p \ d \theta \;,
	\label{eq:adiabatic_inv_def}
\end{equation}
where the integral is over a closed path (\eg a period, $T$), and $C$ indicates that $J$ can always be rescaled with a constant. This quantity is the adiabatic invariant of the system, if the Hamiltonian varies slowly during a cycle. That is,
\[
\dfrac{d J}{d t} = C \ \oint \lrBigb{\dot p \ d \theta + p \ d \dot \theta} = C \ \dint_{t}^{t+T}  \dfrac{\partial \Ham}{\partial t^\prime} \ d t^\prime \approx T \ \dfrac{\partial \Ham(t^{\prime})}{\partial t^{\prime}}\Big|_{t^{\prime}=t} \approx 0 
\;. 
\]

\paragraph{Application to the axion}
The Hamiltonian that results in the EOM of \eqs{eq:eom} is
\begin{equation}
	\Ham = \dfrac{1}{2} \dfrac{p^2}{\fa^2 \ a^3} + V(\theta) \ a^3\;,
	\label{eq:axion_H}
\end{equation}
with 
\begin{eqnarray}
	& p = \fa^2 \ a^3 \ \dot \theta \\
	\label{eq:momentum}
	& V(\theta) = \maT^2 \fa^2 (1-\cos \theta) \;.
	\label{eq:potential}
\end{eqnarray}

\begin{figure}[t]
	\includegraphics[width=1\textwidth]{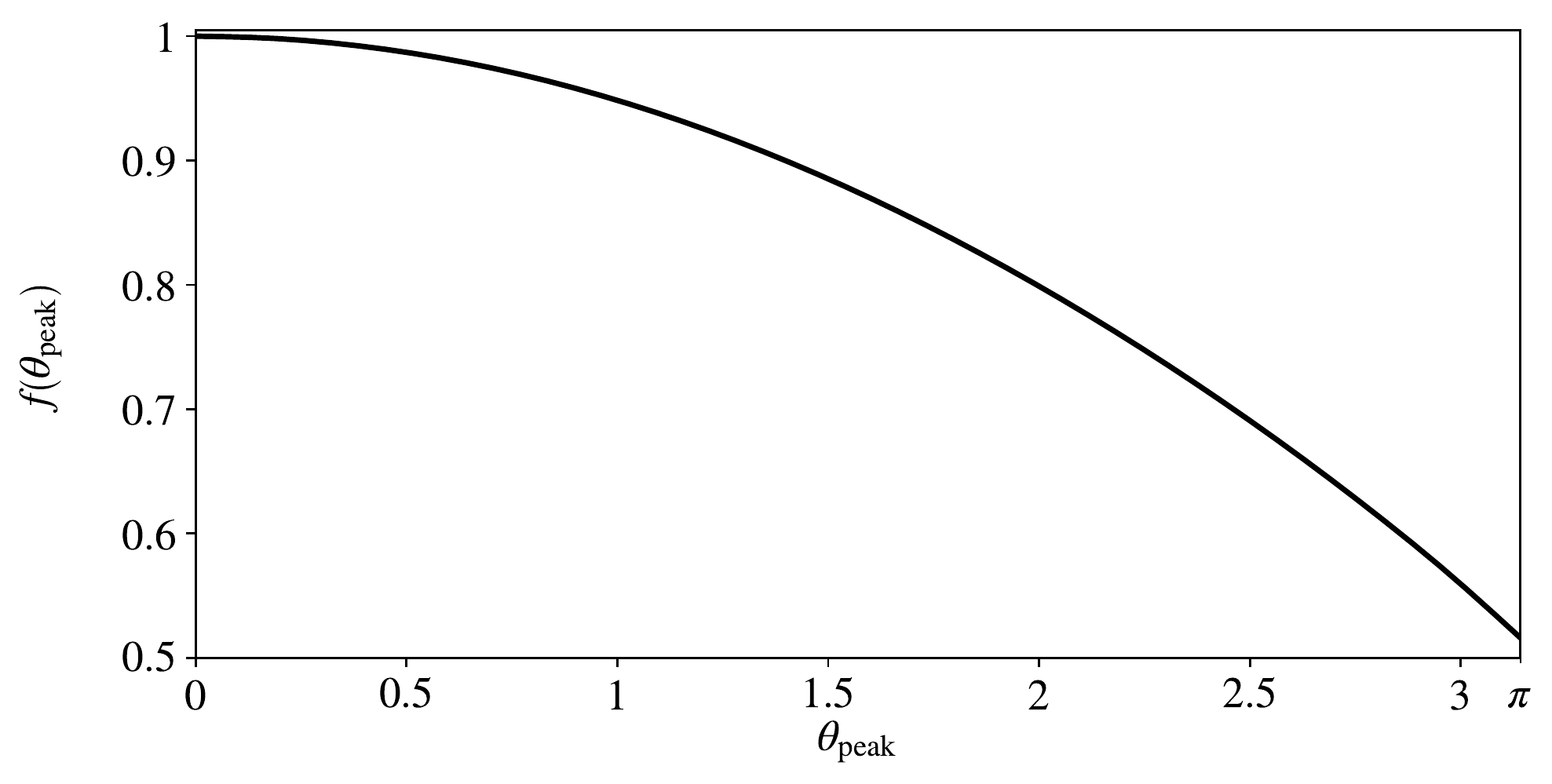}
	\caption{The anharmonic factor for $0 \leq \thetamax < \pi $.}
	\label{fig:anharmonic_factor}
\end{figure}

Notice that the Hamiltonian varies slowly if $\dot {\tilde{m}}_{a}(T)/\maT \ll \maT$ and $H \ll \maT$, which are the adiabatic conditions.  When these conditions are met, the adiabatic invariant for this system becomes
\begin{equation}
	J = \dfrac{\oint p \ d \theta}{\pi \fa^2} = \dfrac{1}{\pi \fa^2} \oint \sqrt{ 2\lrb{ \Ham(\theta) - V(\theta) \ a^3} \ \fa^2 a^3 \ }  \ d \theta  =
	 \dfrac{2}{\pi \fa^2} \int_{-\thetamax}^{\thetamax} \sqrt{ 2\lrb{ \Ham(\thetamax) - V(\theta) \ a^3} \ \fa^2 a^3 \ } d \theta \;,
	 \label{eq:J_axion_definition}
\end{equation}
where we note that $\thetamax$ denotes the maximum of $\theta$ -- the peak of the oscillation, which corresponds to $p=0$. That is, $\Ham(\thetamax) = V(\thetamax) \ a^3$. Therefore, the adiabatic invariant, takes the form 

\begin{eqnarray}
	J=&  \dfrac{2 \sqrt{2} }{\pi \fa}  \int_{- \thetamax} ^{\thetamax}  \sqrt{ V(\thetamax) - V(\theta) } a^{3} d \theta = 
	\dfrac{2 \sqrt{2} }{\pi} \ \maT \, a^3 \ \dint_{- \thetamax}^{\thetamax} \sqrt{\cos \theta - \cos \thetamax} \ d \theta  
	\;,
	\label{eq:J_axion_derivation}
\end{eqnarray}
where, for the last equality. we have used the adiabatic conditions, \ie negligible change of $\maT$ and $a$ during one period. Usually, the adiabatic invariant is written as~\cite{Lyth:1991ub,Bae:2008ue} 
\begin{equation}
	J = a^3 \ \maT \ \thetamax^2  \, f(\thetamax)  \;,
	\label{eq:J_axion_final_form}
\end{equation}
where 
\begin{equation}
	f(\thetamax) =\dfrac{ 2 \sqrt{2}}{\pi \thetamax^2 } \dint_{- \thetamax}^{\thetamax} d \theta \sqrt{ \cos \theta - \cos \thetamax } \;,
	\label{eq:anharmonic_f}
\end{equation}
is called the anharmonic factor, with $ 0.5 \lesssim f(\thetamax) \leq 1$ (see \Figs{fig:anharmonic_factor}).

\paragraph{The role of the adiabatic invariant in the axion relic energy density}
The adiabatic invariant allows us to calculate the maximum value of the angle $\theta$ at late times from its corresponding value at some point just after the adiabatic conditions were fulfilled.

In order to do this, we can numerically integrate \eqs{eq:eom}, and identify the maxima of $\theta$. Once the adiabatic conditions are fulfilled, we can stop the integration at a peak, $\thetamax_{,*}$ -- which corresponds to $T=T_{*}$ and $a=a_{*}$. Then, the value of the maximum angle today ($\thetamax_{,0} \ll 1$) is related to $\thetamax_{,*}$ via
\begin{eqnarray}
	\thetamax_{,0}^2 &=  \lrb{\dfrac{a_*}{a_0}}^3 \ \dfrac{\maT_{,*}}{\ma} \ f(\thetamax_{,*}) \ \thetamax_{,*}^2  =
	\gamma^{-1} \ \dfrac{s_0}{s_*} \ \dfrac{\maT_{,*}}{\ma} \ f(\thetamax_{,*}) \ \thetamax_{,*}^2 
	\; .
	\label{eq:theta_relation}
\end{eqnarray}
Plugging this into \eqs{eq:rho_a_def} (with $\dot \theta=0$, \ie at today's peak), we arrive at the energy density today 
\begin{equation}
	\rho_{a,0} = \gamma^{-1} \ \dfrac{s_0}{s_*} \ \ma \ \maT_{,*} \ \dfrac{1}{2} \ \fa^2 \ \thetamax_{,*}^2 \;  \ f(\thetamax_{,*}) \;,
	\label{eq:rho_axion_exact}
\end{equation}
where $\gamma$ is the entropy injection coefficient between $T_{*}$ and $T_0$, defined from 
\begin{equation}
	a^3(T_0) \ s(T_0) = \gamma \ a^3(T_*) \ s(T_*) \;
	\label{eq:entropy_injection_gamma}
\end{equation}

Notice that \eqs{eq:rho_axion_exact} is similar to the corresponding WKB result~(\ref{eq:rho_a_approx}) at $\tosc \to t_*$, multiplied by the anharmonic factor $f(\thetamax_{,*})$. So, the numerical integration is needed in order to correctly identify $T_{*}$ and $\thetamax_{,*}$, which are greatly affected by the underlying cosmology (especially in cases where entropy injection is active close to $\Tosc$) and whether $\thetai \gtrsim 1$~\cite{Bae:2008ue,Arias:2021rer}. 

It is worth mentioning that \mimes identifies the maxima in real time. Then, integration stops as soon as $J$ becomes almost constant. That is, adiabaticity is assumed to be reached once $J$ does not change significantly between consecutive peaks. \mimes expects from the user to decide  for  how many peaks and how little $J$ needs to change, in order for the system to be considered adiabatic. More details are given in the next section, where we discuss hw \mimes is used.

\section{\mimes usage}\label{sec:first_steps}
\setcounter{equation}{0}
The latest stable version of \mimes is available at \href{https://github.com/dkaramit/MiMeS/tree/stable}{\tt github.com/dkaramit/MiMeS/tree/stable}, which can also be obtained by running:~\footnote{Instructions on how {\tt git} can be installed can be found in \href{https://github.com/git-guides/install-git}{https://github.com/git-guides/install-git}. }
\begin{lstlisting}
	git clone -b stable https://github.com/dkaramit/MiMeS.git
\end{lstlisting}
Moreover, one can download \mimes from \href{https://mimes.hepforge.org}{\tt mimes.hepforge.org}, or by running  
\begin{lstlisting}
	wget -c https://mimes.hepforge.org/downloads/MiMeS-x.y.z.tar.gz -O - | tar -xz
\end{lstlisting}
where {\tt x.y.z} define the version of the library (currently this is {\tt 1.0.0}).

It is important to note that \mimes relies on {\tt NaBBODES}~\cite{NaBBODES} and {\tt SimpleSplines}~\cite{SimpleSplines}. These are two libraries developed independently, and in order to get \mimes with the latest version of these libraries, one needs to run following commands
\begin{lstlisting}
	git clone https://github.com/dkaramit/MiMeS.git
	cd MiMeS
	git submodule init
	git submodule update --remote
\end{lstlisting}
This downloads the {\tt master} branch of \mimes; and {\tt NaBBODES}~\cite{NaBBODES} and {\tt SimpleSplines}~\cite{SimpleSplines} as submodules. This guaranties that \mimes uses the most updated version of these libraries, although it may not be stable.

Once everything is downloaded successfully, we can go inside the \mimes directory, and run \run{bash configure.sh} and \run{make}.  The {\tt bash} script {\tt configure.sh}, just writes some paths to some files, formats the data files provided in an acceptable format (in section~\ref{sec:input} the format is explained), and makes some directories.
The {\tt makefile} is responsible for compiling some examples and checks, as well as the shared libraries that needed for the \PY interface.  If everything runs successfully, there should be two new directories {\tt exec} and {\tt lib}. Inside {\tt exec}, there are several executables that ready to run, in order to ensure that the code runs (\eg no segmentation fault occurs). For example, {\tt \mimes/exec/AxionSolve\_check.run}, should print the values of the parameters $\thetai$ and $\fa$, the oscillation temperature and the corresponding value of $\theta$, the evolution of the axion (\eg temperature, $\theta$, $\rho_{a}$, etc.), and the values of various quantities on the peaks of the oscillation.  In the directory {\tt lib}, there are several shared libraries for the \PY interface.

Although there are various options available at compile-time, we first discuss how \mimes can be used, in order for the role of these options to be clear.

\subsection{First steps}\label{sec:First_examples} 
There are several examples in \CPP ({\tt MiMeS/UserSpace/Cpp}) and \PY ({\tt MiMeS/UserSpace/Python}), as well as \JUPY  notebooks ({\tt MiMeS/UserSpace/JupyterNotebooks}), that show in detail how \mimes can be used. Here, we discuss the various functions one may use as it will provide some insight for the following discussions. 

\subsubsection{Using \mimes in \CPP}\label{sec:cpp_first_example}
The class that is responsible for the solution of the EOM is \cppin{mimes::Axion<LD,Solver,Method>}, located in {\tt MiMeS/src/Axion/AxionSolve.hpp}. However, in order to use it, we first have to define the mass of the axion as a function of the temperature and $\fa$. The axion mass is defined as an instance of the \cppin{mimes::AxionMass<LD>}, which is defined in the header file {\tt \mimes/src/AxionMass/AxionMass.hpp}. We should note that \cppin{LD} is the numerical type to be used (\eg { \tt long double}). The other template arguments are related to the differential equation solver, and their role will be explained in later sections. 

In order to start, at the top of the main {\tt .cpp} file, we need to include 
\begin{cpp}
	#include "src/AxionMass/AxionMass.hpp"
	#include "src/Axion/AxionSolve.hpp"
\end{cpp}
Notice that if the this {\tt .cpp} file is not in the root directory of \mimes, we need to compile it using the flag {\tt -Ipath-to-root}, "path-to-root" the relative path to the root directory of \mimes; \eg if the {\tt .cpp} is in the {\tt MiMeS/UserSpace/Cpp/Axion} directory, this flag should be {\tt -I../../../}.

The mass of the axion can be defined in two ways; either via a data file or as a user defined function. 

\paragraph{Axion mass form data file} 
In many cases, axion mass cannot be written in a closed form. In these cases, the mass is assumed to be of the form of \eqs{eq:axion_mass_def}, and the user has to provide a data file that tabulates the function $\chi(T)$. Then, the axion mass can be defined as
\begin{cpp}
	mimes::AxionMass<LD> axionMass(chi_PATH, minT, maxT);
\end{cpp}
The template parameter \cppin{LD} is a numeric type (\eg \cppin{double} or \cppin{long double}). The argument  {\tt chi\_PATH} is a (relative or absolute) path to a file with two columns; $T$ (in $\GeV$) and $\chi$ (in $\GeV^4$), with increasing $T$. In this case, the axion mass is interpolated between the temperatures {\tt minT} and {\tt maxT}. These two parameters are just suggestions, and the actual interpolation limits, {\tt  TMin} and {\tt TMax}, are chosen as the closest temperatures in the data file. That is, in general, {\tt TMin} $\geq$ {\tt minT} and {\tt TMax} $\leq$ {\tt maxT}. Beyond these limits, the mass is assumed to constant by default. However, one can change this by using 
\begin{cpp}
	axionMass.set_ma2_MAX(ma2_MAX);
	axionMass.set_ma2_MIN(ma2_MIN);
\end{cpp}
with \cppin{ma2_MAX} and \cppin{ma2_MIN} the axion mass squared as functions (or any other callable objects) with signatures \cppin{LD ma2_MAX(LD T, LD fa)} and \cppin{LD ma2_MIN(LD T, LD fa)}. These functions are called for $T\geq${\tt TMax} and $T\leq${\tt TMin}, respectively. Usually, in order to ensure continuity of the axion mass, one needs to know {\tt TMin}, {\tt TMax}, $\chi(${\tt TMin}$)$, and $\chi(${\tt TMax}$)$; which can be found by calling {\tt axionMass.getTMin()}, {\tt axionMass.getTMax()}, {\tt axionMass.getChiMin()}, and {\tt axionMass.getChiMax()}, respectively.

\paragraph{Axion mass from a function} 
In some cases, the dependence of the axion mass on the temperature can be expressed analytically. If this is the case, the user can define the axion mass via 
\begin{cpp}
	mimes::AxionMass<LD> axionMass(ma2);
\end{cpp}
with {\tt ma2} the axion mass squared as a callable object with signature \cppin{LD ma2(LD T, LD fa)}.

\paragraph{The EOM solver}
Once the axion mass is defined, we can declare a variable that will be used to solve the EOM, as
\begin{cpp}
	mimes::Axion<LD,Solver,Method> ax(theta_i, fa, umax, TSTOP, ratio_ini, 
		N_convergence_max, convergence_lim, inputFile, &axionMass, initial_step_size,
		minimum_step_size, maximum_step_size, absolute_tolerance, relative_tolerance, 
		beta, fac_max, fac_min, maximum_No_steps);
\end{cpp}
Here, \cppin{LD} should be the numeric type to be used; it is recommended to use \cppin{long double}, but other choices are also available as we discuss later. Moreover \cppin{Solver} and \cppin{Method} depend on the type of Runge-Kutta (RK) the user chooses. The available choices are shown in table\ref{tab:template-arguments}. 

The various parameters are as follows:
\begin{enumerate}
	\item {\tt theta\_i}: Initial angle.
	\item {\tt fa}: The PQ scale.
	\item {\tt umax }: If $u>${\tt umax} the integration stops (remember that $u=\log(a/a_i)$). Typically, this should be a large number ($\sim 1000$), in order to avoid stopping the integration before the axion begins to evolve  adiabatically.    
	\item {\tt TSTOP}: If the temperature drops below this, integration stops. In most cases this should be around 
	$10^{-4}~\GeV$, in order to be sure that any entropy injection has stopped before integration stops (since BBN bounds~\cite{Kolb:206230,Peebles:1993} should not be violated).
	\item {\tt ratio\_ini}: Integration starts when $3H/\maT \approx${\tt ratio\_ini} (the exact point depends on the file ``{\tt inputFile}", which we will see later). 
	\item  {\tt N\_convergence\_max} and {\tt convergence\_lim}: Integration stops when the relative difference 
	between two consecutive peaks is less than {\tt convergence\_lim} for {\tt N\_convergence\_max} 
	consecutive peaks. This is the point beyond which adiabatic evolution is assumed.
	
	\item  {\tt inputFile}: Relative (or absolute) path to a file that describes the cosmology. the columns should be: $u$ $T ~[\GeV]$ $\log H$, sorted so that $u$ increases.~\footnote{One can run \run{bash MiMeS/src/FormatFile.sh inputFile} in order to sort it and remove any unwanted duplicates. See Appendix~\ref{app:util} for details of {\tt MiMeS/src/FormatFile.sh}.}
	It is important to remember that \mimes assumes that the entropy injection has stopped before the lowest temperature given in {\tt inputFile}. Since \mimes is unable to guess the cosmology beyond what is given in this file, the user has to make sure that there are data between the initial temperature (which corresponds to {\tt ratio\_ini}), and {\tt TSTOP}.
	
	\item {\tt axionMass}: An instance of the \cppin{mimes::AxionMass<LD>} class, passed by pointer. 
	
	\item {\tt initial\_stepsize} (optional): Initial step the solver takes. 
	
	\item {\tt maximum\_stepsize} (optional): This limits the step-size to an upper limit. 
	\item {\tt minimum\_stepsize} (optional): This limits the step-size to a lower limit. 
	
	\item {\tt absolute\_tolerance} (optional): Absolute tolerance of the RK solver
	
	\item {\tt relative\_tolerance} (optional): Relative tolerance of the RK solver.
	Generally, both absolute and relative tolerances should be $10^{-8}$. 
	In some cases, however, one may need more accurate result (\eg if {\tt f\_a} is extremely high, 
	the oscillations happen violently, and the system destabilizes). In any case, if the  
	tolerances are below $10^{-8}$, \cppin{LD} should be \cppin{long double}. \mimes by default uses \cppin{long double} variables, 
	in order to change it see the options available in section~\ref{sec:input}.
	
	\item {\tt beta} (optional): Controls how agreesive the adaptation is. Generally, it should be around but less than 1.
	
	\item {\tt fac\_max},  {\tt fac\_min} (optional): The stepsize does not increase more than fac\_max, and less than fac\_min. 
	This ensures a better stability. Ideally, {\tt fac\_max}$=\infty$ and {\tt fac\_min}$=0$, but in reality one must 
	tweak them in order to avoid instabilities.
	
	\item {\tt maximum\_No\_steps} (optional): Maximum steps the solver can take. Quits if this number is reached even if integration
	is not finished. 
\end{enumerate}
In order to understand the role of the optional parameters, some basic techniques of RK methods are discussed in Appendix~\ref{app:RK}. 

The EOM~(\ref{eq:eom_u}), then can be solved using 
\begin{cpp}
	ax.solveAxion();
\end{cpp}
Once the EOM is solved, we can access $\Tosc$, $\thetaosc$, and $\Omega h^2$  via {\tt ax.T\_osc}, {\tt ax.theta\_osc}, and {\tt ax.relic}. The entire evolution (the points the integrator took) of the axion angle is stored in {\tt ax.points}, which is a two-dimensional \cppin{std::vector<LD>}, with the columns corresponding to  $a$, $T~[\GeV]$, 
$\theta$, $d\theta/du$, $\rho_a$. Moreover, the peaks of he oscillation are stored in another two-dimensional \cppin{std::vector<LD>}, with the columns corresponding to $a$, $T~[\GeV]$, $\thetamax$, $d\theta/du=0$, $\rho_a$, $J$. We should note that the peaks are identified using linear interpolation between integration points, in order to ensure that $d\theta/du = 0$. That is, the values stored in {\tt ax.peaks} do not exist in {\tt ax.points}. 

As already mentioned, \mimes uses embedded RK methods in order to solve the axion EOM. That is, each integration point is calculated twice using two estimates of different order. The difference between the two estimates is then interpreted as the local integration error (see Appendix~\ref{app:RK} for details). These local errors at the integration points of $\theta$ and $\zeta$ are stored in {\tt ax.dtheta} and {\tt ax.dzeta}.

\paragraph{Changing axion mass definition}
The axion mass definiton can change at any time by changing the data file (or {\tt ma2\_MIN} and {\tt ma2\_MAX}) or using 
\begin{cpp}
	ax.set_ma2(ma2);
\end{cpp}
with {\tt ma2} a callable object with signature \cppin{LD ma2(LD T, LD fa)}.

However, since integration starts at a temperature determined by {\tt ratio\_ini}, if the mass changes (including the definitions beyond the interpolation limits), we have to remake the interpolation of the underlying cosmology described by the parameter {\tt inputFile}. Thus, if the definition of the mass changes, we need to call 
\begin{cpp}
	ax.restart();
\end{cpp}
This function remakes the interpolations, clears all vector, and sets all variables to $0$.

\paragraph{Changing initial condition}
The final member function is \cppin{mimes::Axion::setTheta_i}, which allows the user to set a different $\thetai$ without generating another instance.~\footnote{Since the interpolations of the data of {\tt inputFile} are made inside the constructor of the \cppin{mimes::Axion<LD,Solver,Method>} class, \cppin{mimes::Axion<LD,Solver,Method>::setTheta\_i} is a faster choice if ones needs to solve the EOM for a different initial condition.} This function is used as    
\begin{cpp}
	ax.setTeta_i(new_theta_ini);
\end{cpp}
where {\tt new\_theta\_ini} is the new value of $\thetai$. Running this function resets all variables to $0$ (except {\tt T\_osc} and {\tt a\_osc}, since they should not change), and clears all \cppin{std::vector<LD>} variables, which allows the user to simply run \run{ax.solveAxion();} as if {\tt ax} was a freshly defined instance.

\subsubsection{Using \mimes in \PY}\label{sec:begin_py}
The modules for the \PY interface are located in {\tt MiMeS/src/interfacePy}. Although the usage of the {\tt AxionMass} and {\tt Axion} classes is similar to the \CPP case, it is worth showing explicitly how the \PY interface works. One should keep in mind that the various template arguments discussed in the \CPP case have to be chosen at compile-time. That is, for the \PY interface, one needs to choose the numeric type, and RK method to be used when the shared libraries are compiled. This is done by assigning the relevant variable in {\mimes/Definitions.mk} before running \run{make}. The various options are discussed in section~\ref{sec:options}, and outlined in table~\ref{tab:compile_time-options}.

The two relevant classes are defined in the modules {\tt interfacePy.AxionMass} and {\tt interfacePy.Axion}, and can be loaded in a \PY script as 
\begin{py}
	from sys import path as sysPath
	sysPath.append('path_to_src')
	from interfacePy.AxionMass import AxionMass
	from interfacePy.Axion import Axion
\end{py}
It is important that {\tt 'path\_to\_src'} provides the relative path to the {\tt MiMeS/src} directory. For example, if the script is located in {\tt MiMeS/UserSpace/Python}, {\tt 'path\_to\_src'} should be {\tt '../../src'}.

\paragraph{Axion mass definition via a data file}
As before, we first need to define the axion mass. In order to define the axion mass via a file, we use
\begin{py}
	AxionMass axionMass(chi_PATH, minT, maxT)
\end{py}
Here, the constructor requires the same parameters as in \CPP. Moreover, the axion mass beyond the interpolation limits can be changed via
\begin{py}
	axionMass.set_ma2_MAX(ma2_MAX)
	axionMass.set_ma2_MIN(ma2_MIN)
\end{py}
Although the naming is the same as in the \CPP case, there is an important difference. Namely, {\tt ma2\_MAX} and {\tt ma2\_MIN} have to be functions (that take $T$ and $\fa$ as arguments and return $\maT^2$), and cannot be any other callable object. The reason is that \mimes uses the {\tt ctypes} module, which only works with objects compatible with {\tt C}. 
Moreover, the values of {\tt TMin}, {\tt TMax}, $\chi(${\tt TMin}$)$, and $\chi(${\tt TMax}$)$ can be obtained by {\tt axionMass.getTMin()}, {\tt axionMass.getTMax()}, {\tt axionMass.getChiMin()}, and {\tt axionMass.getChiMax()}, respectively.

\paragraph{Axion mass definition via a function}
Again this can be done as
\begin{py}
	AxionMass axionMass(ma2)
\end{py}
with {\tt ma2} the axion mass squared, which should be a function (not any callable object) of $T$ and $\fa$.

\paragraph{Importand note} Once an \pyin{AxionMass} is no longer required, the destructor must be called. In this case, we can run
\begin{py}
	del axionMass
\end{py}
The reason is that \mimes constructs a pointer for every instance of the class, which needs to be deleted manually.  

\paragraph{The EOM solver}
We can define an {\tt Axion} instance as follows 
\begin{py}
	ax=Axion(theta_i, fa, umax, TSTOP, ratio_ini, N_convergence_max, convergence_lim, 
		inputFile, axionMass, initial_step_size, minimum_step_size, maximum_step_size, 
		absolute_tolerance, relative_tolerance, beta, fac_max, fac_min, maximum_No_steps)
\end{py}
Here the input parameters are the same as in the \CPP case, and outlined in table~\ref{tab:AxionSolve-input}. Moreover, the usage of the class can be found by running \pyin{?Axion} after loading the module. The only slight difference compared to the \CPP case is that the \pyin{axionMass} instance is not passed as a pointer; it is done internally using {\tt ctypes}.

Using the defined variable ({\tt ax} in this example), we can simply run  
\begin{py}
	ax.solveAxion()
\end{py}
in order to solve the EOM of the axion. In contrast to the \CPP implementation, this only gives us access to $\Tosc$, $\thetaosc$, and $\Omega h^2$; the corresponding variables are {\tt ax.T\_osc}, {\tt ax.theta\_osc}, and {\tt ax.relic}. In order to get the evolution of the axion field, we need to run 
\begin{py}
	ax.getPoints()
\end{py}
This will make \pyin{numpy}~\cite{harris2020array} arrays that contain the scale factor ({\tt ax.a}), temperature ({\tt ax.T}), $\theta$ ({\tt ax.theta}), its derivative with respect to $u$ ({\tt ax.zeta}), and the energy density of the axion ({\tt ax.rho\_axion}).

Moreover, in order to get the various quantities on the peaks of the oscillation, we can run
\begin{py}
	ax.getPeaks()
\end{py}
This makes \pyin{numpy} arrays that contain the scale factor ({\tt ax.a\_peak}), temperature ({\tt ax.T\_peak}), $\theta$ ({\tt ax.theta\_peak}), its derivative with respect to $u$ ({\tt ax.zeta\_peak}, which should be equal to $0$), the energy density of the axion ({\tt ax.rho\_axion\_peak}), and the values of the adiabatic invariant on the peaks ({\tt ax.adiabatic\_invariant}).

Moreover, we can run the following
\begin{py}
	ax.getErrors()
\end{py}
in order to store the local errors (see Appendix~\ref{app:RK}) of $\theta$ and $\zeta$ in {\tt ax.dtheta} and {\tt ax.dzeta}, respectively. 

\paragraph{Changing axion mass definition}
We can change the axion mass by changing the file, {\tt ma2\_MIN} and {\tt ma2\_MAX}, or using 
\begin{cpp}
	ax.set_ma2(ma2)
\end{cpp}
with {\tt ma2} a function that takes $T$ and $\fa$, and returns the $\maT^2$. As in the \CPP case, if the definition of the mass of the axion changes (including the definitions beyond the interpolation limits), we have to call
\begin{cpp}
	ax.restart()
\end{cpp}
in order to remake the interpolation of cosmological quantities and reset the various variables.

\paragraph{Changing the initial condition}
The initial condition $\thetai$ can be changed using 
\begin{py}
	ax.setTeta_i(new_theta_ini)
\end{py}
which is faster than running the constructor again, since all the interpolations are reused. However, running this function, erases all the arrays, and resets all variables to $0$ (except {\tt T\_osc} and {\tt a\_osc}, as they should not change). 

\paragraph{Importand note} The {\tt Axion} class constructs a pointer to an instance of the underlying \cppin{mimes::Axion} class, which has to be manually deleted. Therefore, once {\tt ax} is used it must be deleted, \ie we need to run 
\begin{py}
	del ax
\end{py}
We should note that this must run even if we assign another instance to the same variable {\tt ax}, otherwise we risk running out of memory.

\section{Assumptions and user input}\label{sec:assumptions}
\setcounter{equation}{0}
\subsection{Restrictions}\label{sec:restrictions}
\mimes only makes a few, fairly general, assumptions. If these assumptions are violated, then \mimes might not work as expected.

First of all, it is assumed that the axion energy density is always subdominant compared to radiation or any other dominant component of the Universe, and that decays and annihilations of particles have a negligible effect on the axion energy density. 

Moreover, the initial condition -- discussed in detail in section~\ref{sec:Physics} -- is always assumed to be $\theta_{t=\ti} = \thetai$ and $\dot \theta|_{t=\ti}=0$. This cannot change, since it is an integral part of how \mimes finds a suitable starting point. Furthermore, it is also assumed that $3H/\maT$ increases monotonically at high temperatures.~\footnote{This assumption is connected to the initial condition, as a sizeable mass at high temperatures can break the shift symmetry mentioned in section~\ref{sec:Physics}.} 

Also, it is assumed that the entropy of the plasma resumes its conserved status at a temperature higher than the minimum temperature of {\tt inputFile} (which is required by the constructor of the \cppin{mimes::Axion<LD,Solver,Method>} class).  

Finally, \mimes does not try to predict anything regarding the cosmology. Therefore, the temperatures in {\tt inputFile} {\em must} cover the entire region of  integration; \ie the maximum temperature has to be larger than the one required to reach {\tt ratio\_ini}, while the minimum one should be lower than {\tt TSTOP}.

\subsection{Options at Compile-time}\label{sec:options}
The user has a number of options regarding different aspects for the code. If \mimes is used without using the available makefiles, then they must use the correct values for the various template arguments, explained in Appendix~\ref{app:classes}.  The various choices we for the shared libraries used by the \PY interface are given in {\tt \mimes/Definitions.mk} while the corresponding options for the \CPP examples are in the {\tt Definitions.mk} files inside the subdirectories of {\tt \mimes/UserSpace/Cpp}. The options correspond to different variables, which are
\begin{enumerate}
	\item {\tt rootDir}: The relative path of root directory of \mimes.  
	\item {\tt LONG}: This sets the numeric types for the \CPP examples. It should be either {\tt long} or omitted. If omitted, the type of the numeric values is \cppin{double} (double precision). On the other hand, if {\tt LONG=long},  the type is  \cppin{long double}. Generally, using \cppin{double} should be enough. For the sake of numerical stability, however, it is advised to always use {\tt LONG=long}, as it a safer option. The reason is that the axion angle redshifts, and can become very small, which introduces ``rounding errors". Moreover, if the parameters {\tt absolute\_tolerance} or {\tt absolute\_tolerance} are chosen to be below $\sim 10^{-8}$, then double precision numbers may not be enough, and {\tt LONG=long} is preferable.  This choice comes at the cost of speed; double precision operations are usually preformed much faster. It is important to note that {\tt LONG} defines a macro with the same name (in the \CPP examples), which then defines the macro (again in the \CPP examples) as {\tt \#define LD LONG double}. The macro \cppin{LD}, then is used as the corresponding template argument in the various classes. We point out again that if one chooses not to use the {\tt makefile} files, the template arguments need to be known at compile-time. So the user has to define them in the code. 
	\item {\tt LONGpy}: the same as {\tt LONG}, but for the \PY interface. One should keep in mind thtat this cannot be changed inside \PY scripts. It just instructs \pyin{ctypes} what numeric type to use. Since the preferred way to compile the shared libraries is via running \run{make} in the root directory of \mimes, this variable needs to be defined inside  {\tt \mimes/Definitions.mk}. By default, this variable is set to {\tt long}, since this is the most stable choice in general.  
	\item {\tt SOLVER}: \mimes uses the ordinary differential equation (ODE) integrators of ref.~\cite{NaBBODES}. Currently, there are two available choices; {\tt Rosenbrock} and {\tt RKF}. The former is a general embedded Rosenbrock implementation and it is used if {\tt SOLVER=1}, while the latter is a general explicit embedded Runge-Kutta implementation and can be chosen by using {\tt SOLVER=2} (a brief description of how these algorithms are implemented can be found in Appendix~\ref{app:RK}). By default inside the {\tt Definitions.mk} files {\tt SOLVER=1}, because the axion EOM tends to oscillate rapidly. However, in some cases, a high order explicit method may also work. Note that this variable defines a macro that is then used as the second template argument of the \cppin{mimes::Axion<LD,Solver,Method>} class. The preferred way to do it in the shared libraries is via the {\tt \mimes/Definitions.mk} file, however, the user if free to compile everything in a different way. In this case, the  various {\tt Definitions.mk} files, are not being used, and the user must define the relevant arguments in the code where \mimes is used.
	\item {\tt METHOD}: Depending on the type of solver, there are some available methods.~\footnote{It is worth mentioning that {\tt NaBBODES} is built in order to be a template for all possible Rosenbrock and explicit Runge-Kutta embedded methods, and one can provide their own Butcher tableau if they want to use another method, as shown in Appendix~\ref{app:RK}.}  
	\begin{itemize}
		\item 	For {\tt SOLVER=1}, the available methods are 
		{\tt METHOD=RODASPR2} and {\tt METHOD=ROS34PW2}. The {\tt RODASPR2} choice is a fourth order Rosenbrock-Wanner method (more information can be found in ref.~\cite{RANG2015128}). The {ROS34PW2} choice corresponds to a third order Rosenbrock-Wanner method~\cite{RangAngermann2005}. 
		\item 	For {\tt SOLVER=2}, the only reliable method available in {\tt NaBBODES} is the Dormand-Prince~\cite{DORMAND198019} chosen if {\tt METHOD=DormandPrince}, which is an explicit Runge-Kutta method of seventh order.
	\end{itemize}
	This variable defines a macro (with the same name) that is passed as the third template parameter of \cppin{mimes::Axion<LD,Solver,Method>} (\ie \cppin{METHOD<LD>} in the place of \cppin{Method}). 
	If the compilation is not done via the {\tt makefile} files, the user must define the relevant template arguments in the code.
	\item {\tt CC}: The \CPP compiler that one chooses to use. The default option is {\tt CC=g++}, which is the {\tt GNU} \CPP compiler, and is available for all systems. Another option is to use the {\tt clang} compiler, which is chosen by {\tt CC=clang -lstdc++}. \mimes is mostly tested using {\tt g++}, but {\tt clang} also seems to work (and the resulting executables are sometimes faster), but the user has to make sure that their version of the compiler of choice supports the \CPP17 standard, otherwise \mimes probably will not work.
	\item {\tt OPT}: Optimization level of the compiler. By default, this is {\tt OPT=O3}, which produces executables that are marginally faster than {\tt OPT=O1} and {\tt OPT=O2}, but significantly faster than {\tt OPT=O0}. There is another choice, {\tt OPT=Ofast}, but it can cause various numerical instabilities, and is generally considered dangerous -- although we have not observed any problems when running \mimes. 
\end{enumerate}
It is important to note, once again, that the variables that correspond to template arguments must be known at compile time. Thus, if the compilation is done without the help of the various {\tt makefile} files, the template arguments must be given, otherwise compilation will fail.~\footnote{In \CPP the template arguments are part of the definition of a class; if the template arguments are not known, the class is not even constructed.} For example, the choice {\tt LONG=long}, {\tt SOLVER=1}, and {\tt METHOD=RODASPR2} will be used to compile the shared libraries (and \CPP example in {\tt \mimes/UserSpace/Cpp/Axion}) with \cppin{mimes::Axion<long double,1,RODASPR2<long double>>}. In order to fully understand the template arguments, the signatures of all classes and functions are given in Appendix~\ref{app:classes}.

\subsection{User input}\label{sec:input}
\subsubsection{Compile-time input}\label{sec:compile_time_input} 
\paragraph{Files} \mimes requires files that provide data for the relativistic degrees of freedom (RDOF) of the plasma, and the anharmonic factor. Although \mimes is shipped with the standard model RDOF found in~\cite{Saikawa:2020swg}, and a few points for $f(\thetamax)$ introduced in \eqs{eq:anharmonic_f}, the user is free change them via the corresponding variables in {\tt \mimes/Paths.mk}. Moreover, there is a set of data for the QCD axion mass as calculated in ref.\cite{Borsanyi:2016ksw}.
The variables pointing to these data files are {\tt cosmoDat}, {\tt axMDat}, and {\tt anFDat}, for the RDOF, axion mass, and the anharmonic factor; respectively.

The format of the files has to be the following:
\begin{itemize}
	\item The RDOF data must be given in three columns; $T$ (in $\GeV$), $\heff$, and $\geff$.
	\item The axion mass data must be given in two columns; $T$ (in $\GeV$), $\chi$ (in $\GeV^4$). Here, $\chi$ is defined as in \eqs{eq:axion_mass_def}. 
	The user can provide a function instead of data for the axion mass, by leaving the  {\tt axMDat} variable empty. 
	\item The data for the anharmonic factor must be give in two columns   $\thetamax$ $f(\thetamax)$; with increasing $\thetamax$.
\end{itemize}
The paths to these files should be given at compile time. That is, once {\tt Paths.mk} changes, we must run \run{bash configure.sh} and then \run{\tt make} in order to make sure that they will be used. The user can change the content of the data files (without changing their paths), in order to use them without compiling \mimes again. However, the user has to make sure that all the files are sorted so that the values of first column increase (with no duplicates or empty lines). In order to ensure this, it is advised to run \run{bash FormatFile.sh path-to-file} (in Appendix~\ref{app:util} there are some details on {\tt MiMeS/src/FormatFile.sh}), in order to format the file (that should exist in \run{path-to-file}) so that it complies with the requirements of \mimes.

These paths are stored as strings in {\tt \mimes/src/misc\_dir/path.hpp} at compile-time (they are defined as {\tt constexpr}), and can be accessed once this header file is included. The corresponding variables are {\tt cosmo\_PATH}, {\tt chi\_PATH}, and {\tt anharmonic\_PATH}, for the path to data file of the plasma quantities, $\chi(T)$, and $f(\thetamax)$; respectively. Although, the axion mass data file may be omitted -- since the axion mass is defined by the user, the variable {\tt chi\_PATH} is still useful if the axion mass is defined via a data file, as it is automatically converted to an absolute path.~\footnote{Absolute paths have the advantage to be accessible from everywhere else in the system. Thus, executables that seek the corresponding files can be called and copied easily.}

\subsubsection{Run-time input}\label{sec:run_time_input}
The run-time user input is described in sec.~\ref{sec:first_steps}. The user has to provide the parameters that describe the axion evolution, $\thetai$ and $\fa$. 

Moreover,  the maximum allowed value of $u$ and the minimum value of $T$, allow the user to decide when  integration stops even if the axion has not reached its adiabatic evolution. Ideally, {\tt umax}$=\infty$ and {\tt TSTOP}$=0$, but \mimes is designed to be as general as possible and there may be cases where one needs to stop the integration abruptly.~\footnote{These two variables are not optional, because the user must be aware of them, in order to choose them according to their needs.}

Furthermore, {\tt ratio\_ini} allows the user to choose a desired point at which the interpolation of the data in {\tt inputFile} begins. This can save valuable computation time as well as memory, as only the necessary data are stored and searched. Generally, for {\tt ratio\_ini}$>10^{3}$, the relic abundance becomes
independent from  {\tt ratio\_ini}, but one has to choose it carefully, in order to find a balance between accuracy and computation time.

Finally, the convergence conditions -- \ie~{\tt N\_convergence\_max} and {\tt convergence\_lim} -- allow the user to decide when the adiabatic evolution of the axion begins. Generally, the relic abundance does not have a strong dependence on these parameters as long as {\tt N\_convergence\_max}$>3$ and {\tt convergence\_lim}$<10^{-1}$ (\ie the adiabatic invariant does not vary more that $10\%$ for three consecutive peaks of the oscillation). 
However, we should note that greedy choices (\eg~{\tt N\_convergence\_max}$=100$ and {\tt convergence\_lim}$=10^{-5}$) are dangerous, as $\theta$ tends to oscillate rapidly which  destabilizes the differential equation solver. Therefore, these parameters should be chosen carefully, in order to ensure that integration stops when the axion has reached its adiabatic evolution, without destabilising the EOM.

\subsection{Complete Examples}\label{sec:complete_examples}
\begin{figure}[t]
	\includegraphics[width=1\textwidth]{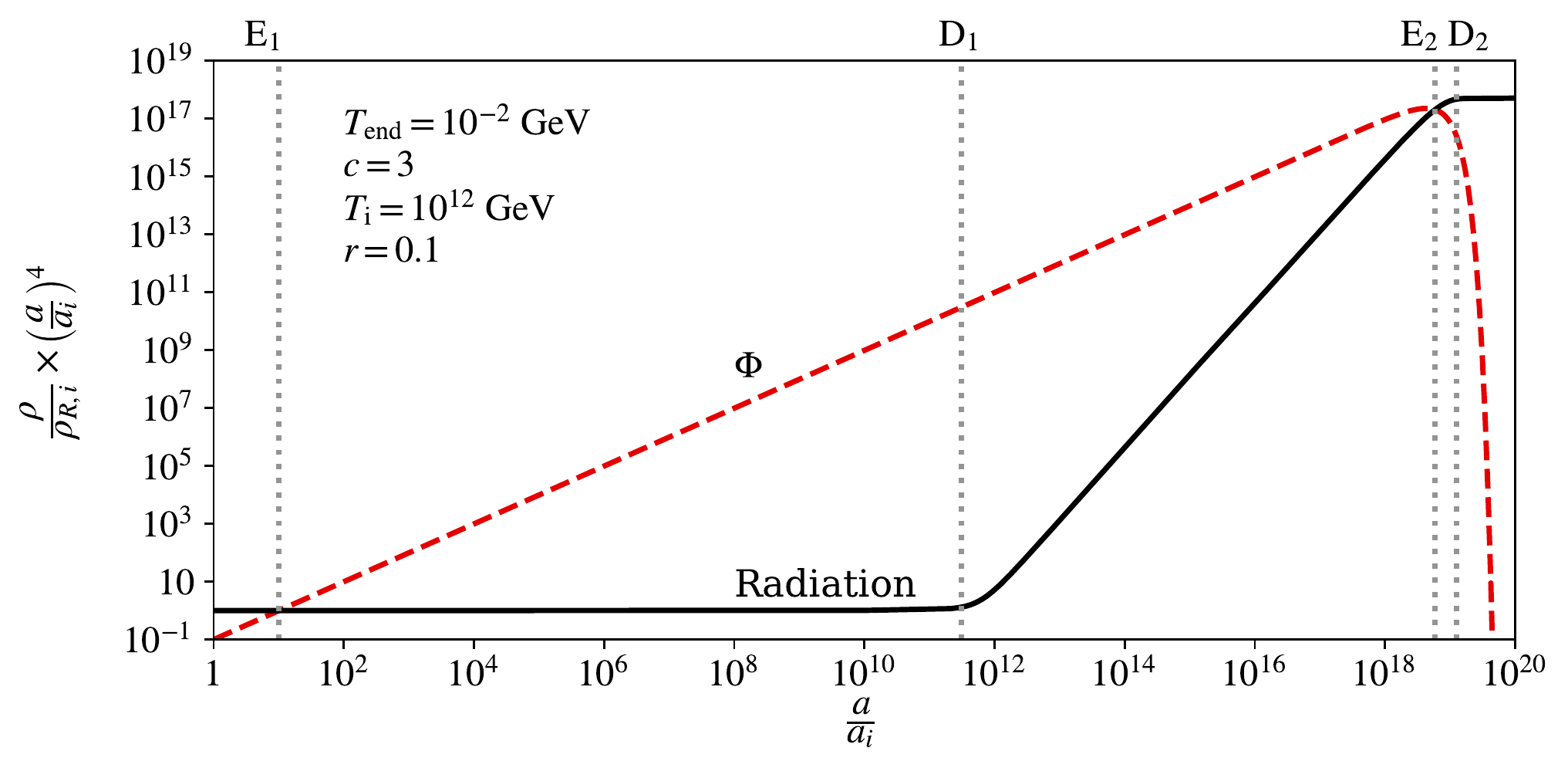}
	\caption{The energy densities of the plasma and a hypothetical decaying fluid. The parameters of the cosmological scenario  $T_{\rm END}=10^{-2}~\GeV$, $c=3$, $T_{\rm ini}=10^{12}~\GeV$, and $r=0.1$; the parameters  follow the definitions in ref.~\cite{Arias:2020qty}. }
	\label{fig:energy_densities}
\end{figure}

Although one can modify the examples provided in {\tt \mimes/UserSpace}, in this section we show a complete example in both \CPP and \PY. 
The underlying cosmology is assumed to be an EMD scenario,~\footnote{In terms of the parametrisation introduced in ref.~\cite{Arias:2019uol,Arias:2020qty},  for this scenario we choose $T_{\rm end}=10^{-2} ~\GeV,\; c=3, \; T_{\rm ini}=10^{12} ~\GeV, \; \text{and}\; r=0.1$.}.
with the evolution of the energy densities of the plasma and the matter field ($\Phi$) shown in \Figs{fig:energy_densities}.
The various regime changes are indicated as $T_{{\rm E}_{1,2}}$ -- corresponding to the first and second time where $\rho_{\Phi} = \rho_{\rm R}$,  and $T_{{\rm D}_{1,2}}$ -- which correspond to the start and end of entropy injection (for more precise definition see~\cite{Arias:2020qty}). Regarding the axion, this cosmological scenario alters its evolution, since both the Hubble parameter and the entropy of the plasma change significantly. This results in a shifted oscillation temperature, and the dilution of the axion energy density.

Moreover, we will use the QCD axion mass of ref.~\cite{Borsanyi:2016ksw}. For the axion mass beyond the minimum and maximum temperatures, $T_{\rm min, \, max}$, in the corresponding data file, we will take
\begin{equation}
	\maT = \Bigg\{
	\begin{matrix}
		\dfrac{\chi(T_{\rm min})}{\fa^2} & \text{for } T<T_{\rm min} 
		\\ \\
		\dfrac{\chi(T_{\rm max})}{\fa^2}   \lrb{\dfrac{T}{T_{\rm max}}}^{-8.16} & \text{for } T>T_{\rm max} 
	\end{matrix} \;.
	\label{eq:axM-limits}
\end{equation}

\subsubsection{complete example in \CPP}\label{sec:cpp_example}
In order to write a \CPP program that uses \mimes in order to solve the EOM~\ref{eq:eom_sys}, we must include the header files {\tt src/AxionMass/AxionMss.hpp} and {\tt src/Axion/AxionSolve.hpp}. In the example at hand, the {\tt main} function should contain the definition of the axion mass
\begin{cpp}
	// use chi_PATH to interpolate the axion mass.
	mimes::AxionMass<long double> axionMass(chi_PATH,0,mimes::Cosmo<long double>::mP);

	/* This is the axion mass squared beyond the interpolation limits 
	for the current data. If you don't specify them, the axion mass 
	is taken to be constant beyond these limits */

	/*set ma2 for T>TMax*/
	long double TMax=axionMass.getTMax();    
	long double chiMax=axionMass.getChiMax();    
	axionMass.set_ma2_MAX(
		[&chiMax,&TMax](long double T, long double fa){
			return chiMax/fa/fa*std::pow(T/TMax,-8.16);
		}
	);  
	
	/*set ma2 for T<TMin*/
	long double TMin=axionMass.getTMin();  
	long double chiMin=axionMass.getChiMin();    
	axionMass.set_ma2_MIN( 
		[&chiMin,&TMin](long double T, long double fa){
			return chiMin/fa/fa;
		}
	);
\end{cpp}
It should be noted that, since we have used {\tt chi\_PATH}, we need to include the {\tt path.hpp} header file from {\tt \mimes/src/misc\_dir}. Moreover, we note that since the second and third arguments are $0$ and \cppin{mimes::Cosmo<LD>::mP} (the Planck mass; see~\ref{app:classes}), interpolation of the axion mass will be between the minimum and maximum values that appear in the data file.

Following, the axion EOM can be defined and solved as
\begin{cpp}
 	std::string inputFile = std::string(rootDir)+
    				std::string("/UserSpace/InputExamples/MatterInput.dat");
	
	mimes::Axion<long double, 1, RODASPR2<long double> > 
					ax(0.1, 1e16, 500, 1e-4, 1e3, 15, 1e-2, inputFile, &axionMAss, 
					1e-1, 1e-8, 1e-1, 1e-11, 1e-11, 0.9, 1.2, 0.8, int(1e7));
					
	ax.solveAxion();
\end{cpp}
In order to print anything, however, we need to write a few more lines of code. For example, the relic abundance is printed by adding \cppin{ std::cout<<ax.relic<<"\n";} after \cppin{ax.solveAxion();}. It should be noted that, in order to make sure that the code is compiled successfully, one should add at the top of the file any other header they need. In particular, in this case, we need to add \cppin{#include<iostream>}, in order to be able to use \cppin{std::cout}.

\paragraph{Parameter choice}
It should be noted that the string that is assigned to the variable {\tt inputFile}, is a path of a file that exists in {\tt \mimes/UserSpace/InputExamples}. Also, {\tt rootDir} is a constant static ({\tt char[]}) variable that  is defined in {\tt \mimes/src/path.hpp}, and it is automatically generated when {\tt bash configure.sh} is executed.  Moreover, the other parameters used in the constructor are:
\begin{itemize}
	\begin{minipage}{0.3\linewidth}
	\item {\tt theta\_i}$=0.1$
	\item {\tt fa}$=10^{16}~\GeV$
	\item {\tt umax }$=500$
	\item {\tt TSTOP}$=10^{-4}~\GeV$
	\item {\tt ratio\_ini}$=10^3$
	\item  {\tt N\_convergence\_max}$=15$
	\end{minipage}
	\begin{minipage}{0.35\linewidth}
	\item  {\tt convergence\_lim}$=10^{-2}$
	\item {\tt initial\_stepsize}$=10^{-1}$ 
	\item {\tt maximum\_stepsize}$=10^{-1}$ 
	\item {\tt minimum\_stepsize}$=10^{-8}$
	\item {\tt absolute\_tolerance}$=10^{-11}$
	\item {\tt relative\_tolerance}$=10^{-11}$
	\end{minipage}
	\begin{minipage}{0.3\linewidth}
	\item {\tt beta}$=0.9$
	\item {\tt fac\_max}$=1.2$
	\item {\tt fac\_min}$=0.8$
	\item {\tt maximum\_No\_steps}$=10^7$
	\item[]{\vfill}	\item[]{\vfill}	\item[]{\vfill}
	\end{minipage}
\end{itemize}

\paragraph{Template arguments}
Furthermore, The first template parameter is \cppin{long double}, which means that all the numeric types (apart from integers like {\tt maximum\_No\_steps}) have ``long double" precision. which is useful when considering low tolerances as in this example. The second and third template parameters are responsible for the Runge-Kutta method we use. That is, in this case, the second template parameter -- $1$ -- means that we use a Rosenbrock method, with the method being {\tt RODASPR2}. Notice that the method also needs a template parameter that is used to declare all member variables of the {\tt RODASPR2} class as \cppin{long double}.

\paragraph{Compilation}
Before we compile, we have to make sure that {\tt bash configure.sh} has been executed. Assuming that we name the file that contains the code is {\tt axionExample.cpp}, and it is located in {\tt \mimes/UserSpace}, an executable can be produced as
\begin{pseudo}
	g++ -O3 -std=c++17 -lm -I../ -o axion axionExample.cpp
\end{pseudo}
or 
\begin{pseudo}
	clang -lstdc++ -O3 -std=c++17 -lm -I../ -o axion axionExample.cpp
\end{pseudo}

Both of these commands should create an executable that solves the axion EOM. That is, assuming we have a terminal open in   {\tt \mimes/UserSpace}, we can run {\tt ./axion}, and get the results we chose. It should be noted that all paths that \mimes uses by default are written as absolute paths in {\tt \mimes/src/misc\_dir/paths.hpp} when we run {\tt bash configure.sh}. Therefore, if the {\tt inputFile} variable is also an absolute path, the executable can be copied and used in any other directory of the same system. However, it should be preferred that executables are kept under the \mimes directory, in order to be able to compile them with different data file paths if needed.

\paragraph{The entire code}
This example consists of only a few lines of code, which, including the change in the axion mass beyond the interpolation limit, is
\begin{cpp}
	#include<iostream>
	//include everything you need from MiMeS
	#include"src/Axion/AxionSolve.hpp"
	#include"src/AxionMass/AxionMass.hpp"
	#include"src/Cosmo/Cosmo.hpp"
	#include"src/misc_dir/path.hpp"
	
	int main(){		
		// use chi_PATH to interpolate the axion mass.
		mimes::AxionMass<long double> axionMass(chi_PATH,0,mimes::Cosmo<long double>::mP);
		
		/*set ?$\maT^2$? for ?$T\geq T_{\rm max}$?*/
		long double TMax=axionMass.getTMax();    
		long double chiMax=axionMass.getChiMax();    
		axionMass.set_ma2_MAX(
			[&chiMax,&TMax](long double T, long double fa){
				return chiMax/fa/fa*std::pow(T/TMax,-8.16);
			}
		);  
		
		/*set ?$\maT^2$? for ?$T\leq T_{\rm min}$?*/
		long double TMin=axionMass.getTMin();  
		long double chiMin=axionMass.getChiMin();    
		axionMass.set_ma2_MIN( 
			[&chiMin,&TMin](long double T, long double fa){
				return chiMin/fa/fa;
			}
		);		

		std::string inputFile = std::string(rootDir)+
			std::string("/UserSpace/InputExamples/MatterInput.dat");
		
		mimes::Axion<long double, 1, RODASPR2<long double> > 
		ax(0.1, 1e16, 500, 1e-4, 1e3, 15, 1e-2, inputFile, &axionMass, 
		1e-1, 1e-8, 1e-1, 1e-11, 1e-11, 0.9, 1.2, 0.8, int(1e7));
		
		ax.solveAxion();
		
		std::cout<<ax.relic<<"\n";
		
		return 0;
	}
\end{cpp}
We should point out that another example is given in {\tt \mimes/UserSpace/Cpp/Axion}, where all parameters are taken as command-line inputs, the various compilation-time options can be given {\tt \mimes/UserSpace/Cpp/Axion/Definitions.mk}, and then compiled using {\tt make}. Therefore, the user can just modify this code in order to meet their needs. That is, using this example, one only needs to add their preferred definition of {\tt ma2\_MAX} and {\tt ma2\_MIN} -- or change the {\tt axionMass} variable (in {\tt \mimes/src/static.hpp}) using a function as mentioned in section~\ref{sec:compile_time_input} without writing the entire program themselves.

\paragraph{Alternative axion mass definition}
For this particular example, we could have used the approximate definition of the axion mass
\begin{equation}
	\maT^2 \approx \ma^2 \times \Bigg\{ 
	\begin{matrix}
	\lrb{\dfrac{T_{\rm QCD}}{T}}^{8.16} & \quad \text{for } T\geq T_{\rm QCD} 
	\\ \\
	1 & \quad \text{for } T\leq T_{\rm QCD}
\end{matrix} \;,
	\label{eq:axM_approx}
\end{equation}
where $T_{\rm QCD}  \approx 150~\MeV$ and $\ma=\dfrac{3.2 \times 10^{-5} ~\GeV^4}{\fa^2}$. This can be done by substituting the {\tt axionMass} definition (lines $9$-$28$) with 
\begin{cpp}
	auto ma2 = [](long double T,long double fa){
	    long double TQCD=150*1e-3;
	    long double ma20=3.1575e-05/fa/fa;
	    if(T<=TQCD){return ma20;}
	    else{return ma20*std::pow((TQCD/T),8.16);}
	};
	mimes::AxionMass<long double> axionMass(ma2);
\end{cpp}

\subsubsection{complete example in \PY}
In order to be able use the \pyin{AxionMass} and \pyin{Axion} classes in \PY, we need to import the corresponding modules from {\tt \mimes/src/interfacePy}. That is, assuming that the script from which we intend to import {\tt \mimes/src/interfacePy/Axion/Axion.py} and {\tt \mimes/src/interfacePy/Axion/AxionMass.py} is in {\tt \mimes/UserSpace}, on top of the script, we need to write
\begin{py}
	#add the relative path for MiMeS/src
	from sys import path as sysPath
	sysPath.append('../src')
	
	from interfacePy.AxionMass import AxionMass #import the AxionMass class
	from interfacePy.Axion import Axion #import the Axion class
	from interfacePy.Cosmo import mP #import the Planck mass
\end{py} 
Once everything we need is imported, we can simply follow the steps outlined in section~\ref{sec:begin_py}. For the example at hand, we can create an instance of the \pyin{AxionMass} class as
\begin{py}
	# AxionMass instance
	axionMass = AxionMass(r'../src/data/chi.dat',0,mP)
	
	# This is the axion mass squared beyond the interpolation limits for the current data. 
	# If you don't specify them, the axion mass is taken to be constant beyond these limits
	TMax=axionMass.getTMax() 
	chiMax=axionMass.getChiMax()
	TMin=axionMass.getTMin() 
	chiMin=axionMass.getChiMin()
	
	axionMass.set_ma2_MAX( lambda T,fa: chiMax/fa/fa*pow(TMax/T,8.16))
	axionMass.set_ma2_MIN( lambda T,fa: chiMin/fa/fa )
\end{py}
Similar to the \CPP case, the second and third arguments are $0$ and \pyin{mP} (the Planck mass). However, in the \PY interface, {\tt mP} needs to be imported from the \pyin{Cosmo} module.

Then we can simply create an instance of the \pyin{Axion} class as 
\begin{py}
	#in python it is more convenient to use relative paths
	inputFile = "./InputExamples/MatterInput.dat"  
	
	ax = Axion(0.1, 1e16, 500, 1e-4, 1e3, 15, 1e-2, inputFile, axionMass, 
					1e-1, 1e-8, 1e-1, 1e-11, 1e-11, 0.9, 1.2, 0.8, int(1e7))
\end{py}
Again, the parameters passed to the constructor are the same as in the \CPP example. The axion EOM, then, is solved using
\begin{py}
	ax.solveAxion()
\end{py}
In contrast to the \CPP usage of this function, this only stores the $\thetai$, $\fa$, $\thetaosc$, $\Tosc$, and $\Omega h^2$ in the variables {\tt ax.theta\_i}, {\tt ax.fa}, {\tt ax.theta\_osc}, {\tt ax.T\_osc}, and {\tt ax.relic}; respectively. Therefore, we can print the relic abundance by calling \pyin{print(ax.relic)}. In order to get the integration points (\ie the evolution of the angle and the other quantities), the quantities at the  peaks, and the local integration errors, we need to call
\begin{py}
	#this gives you all the points of integration
	ax.getPoints()
	
	#this gives you the peaks of the oscillation
	ax.getPeaks()
	
	#this gives you local integration errors
	ax.getErrors()
\end{py}
The documentation of any \PY function can be read directly inside the script using {\tt ?} as a prefix. For example, in order to see what the functionality and usage of the {\tt getPeaks} function, we can call \run{?ax.getPeaks}, and its documentation will be printed. 

As already mentioned, it is important to always delete any instance of the \pyin{AxionMass} and \pyin{Axion} classes once they are not needed. In this case this is done by calling
\begin{py}
	del ax
	del axionMass
\end{py}

\paragraph{Compilation of the shared library}
As described in section~\ref{sec:compile_time_input}, we may need to change the default data file paths, or the various compilation options. This is done through the variables in {\tt \mimes/Definitions.mk} and {\tt \mimes/Paths.mk} described in section~\ref{sec:options}. 

Once we have chosen everything according to our needs, the library can be created by opening a terminal inside the root directory of \mimes and running
\begin{py}
	bash configure.sh
	make lib/Axion_py.so
\end{py}

\paragraph{The entire code}
As in the \CPP example, this example consists of only a few lines of code. The script we described here is
\begin{py}
	#add the relative path for MiMeS/src
	from sys import path as sysPath
	sysPath.append('../src')	
	
	from interfacePy.AxionMass import AxionMass #import the AxionMass class
	from interfacePy.Axion import Axion #import the Axion class
	from interfacePy.Cosmo import mP #import the Planck mass

	# AxionMass instance
	axionMass = AxionMass(r'../src/data/chi.dat',0,mP)
	
	# define ?$\maT^2$? for ?$T\leq T_{\rm min}$?
	TMin=axionMass.getTMin() 
	chiMin=axionMass.getChiMin()
	axionMass.set_ma2_MIN( lambda T,fa: chiMin/fa/fa )

	# define ?$\maT^2$? for ?$T\geq T_{\rm max}$?
	TMax=axionMass.getTMax() 
	chiMax=axionMass.getChiMax()
	axionMass.set_ma2_MAX( lambda T,fa: chiMax/fa/fa*pow(TMax/T,8.16))
	
	#in python it is more convenient to use relative paths
	inputFile = inputFile="./InputExamples/MatterInput.dat"  
	
	ax = Axion(0.1, 1e16, 500, 1e-4, 1e3, 15, 1e-2, inputFile, axionMass, 
					1e-1, 1e-8, 1e-1, 1e-11, 1e-11, 0.9, 1.2, 0.8, int(1e7))

	ax.solveAxion()
	
	print(ax.relic)
	
	#once we are done we should run the destructor
	del ax
	del axionMass
\end{py}

One can find a complete example, including the option to create several plots, in the script {\tt \mimes/UserSpace/Python/Axion.py}. Also, the same example can be used interactively, in jupyter notebook environment~\cite{Kluyver2016jupyter}, that can be found in {\tt \mimes/UserSpace/JupyterNotebooks/Axion.ipynb}. One can read the comments, and change all different parameters, in order to examine how the results are affected. 

\paragraph{Alternative axion mass definition}
In order to use the approximate axion mass as defined in \eqs{eq:axM_approx}, we could replace the definition of the {\tt axionMass} variable (lines $10$-$21$) with
\begin{py}
	def ma2(T,fa):
		TQCD=150*1e-3
		ma20=3.1575e-05/fa/fa
		if T<=TQCD:
			return ma20;
		return ma20*(TQCD/T)**8.16
	axionMass = AxionMass(ma2)
\end{py}

\subsubsection{Results}
\begin{figure}[h]
	\begin{subfigure}[]{0.5\textwidth}
		\includegraphics[width=1\textwidth]{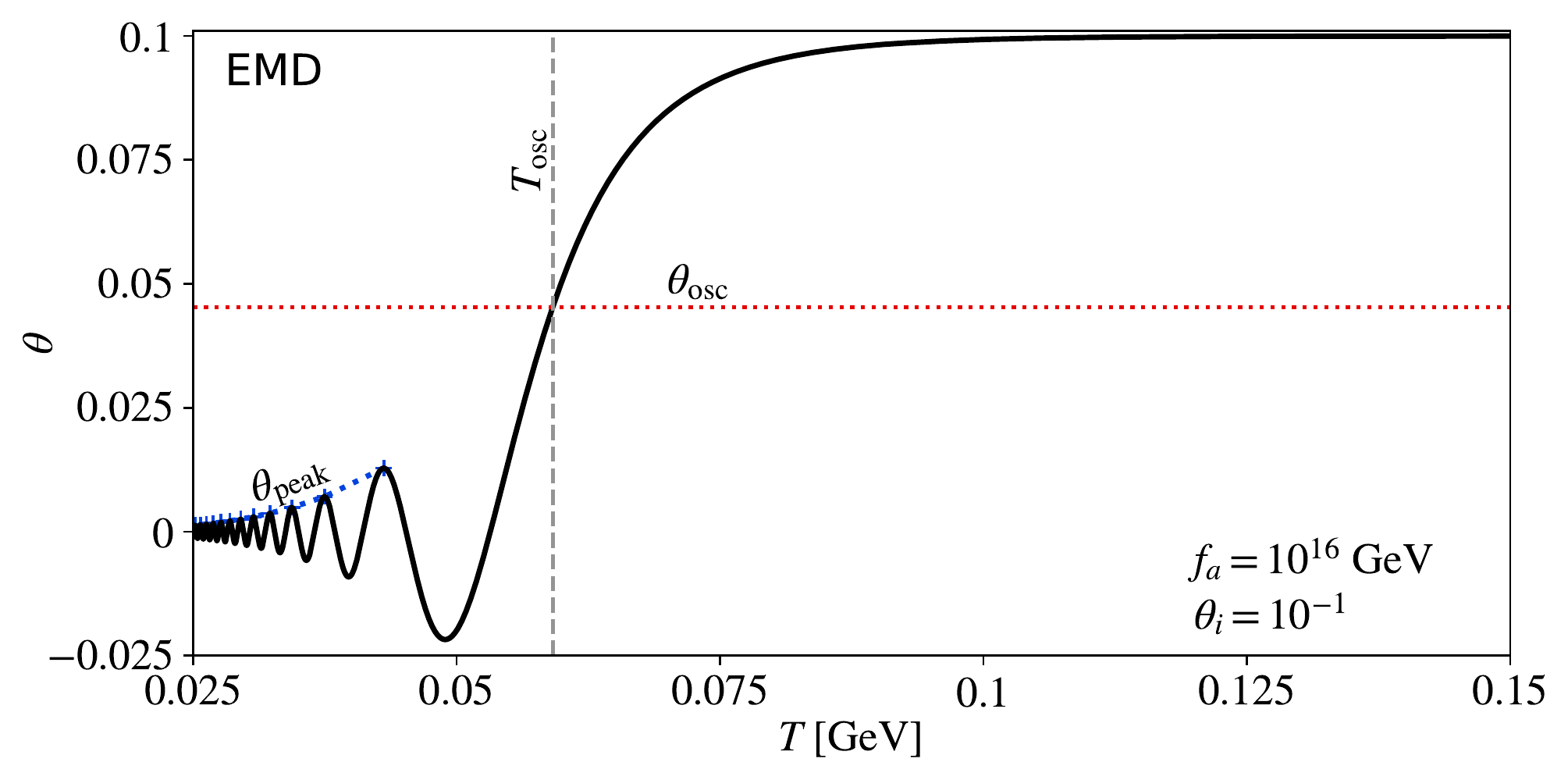}
		\caption{}
		\label{fig:theta_evolution-EMD}
	\end{subfigure}
	\begin{subfigure}[]{0.5\textwidth}
		\includegraphics[width=1\textwidth]{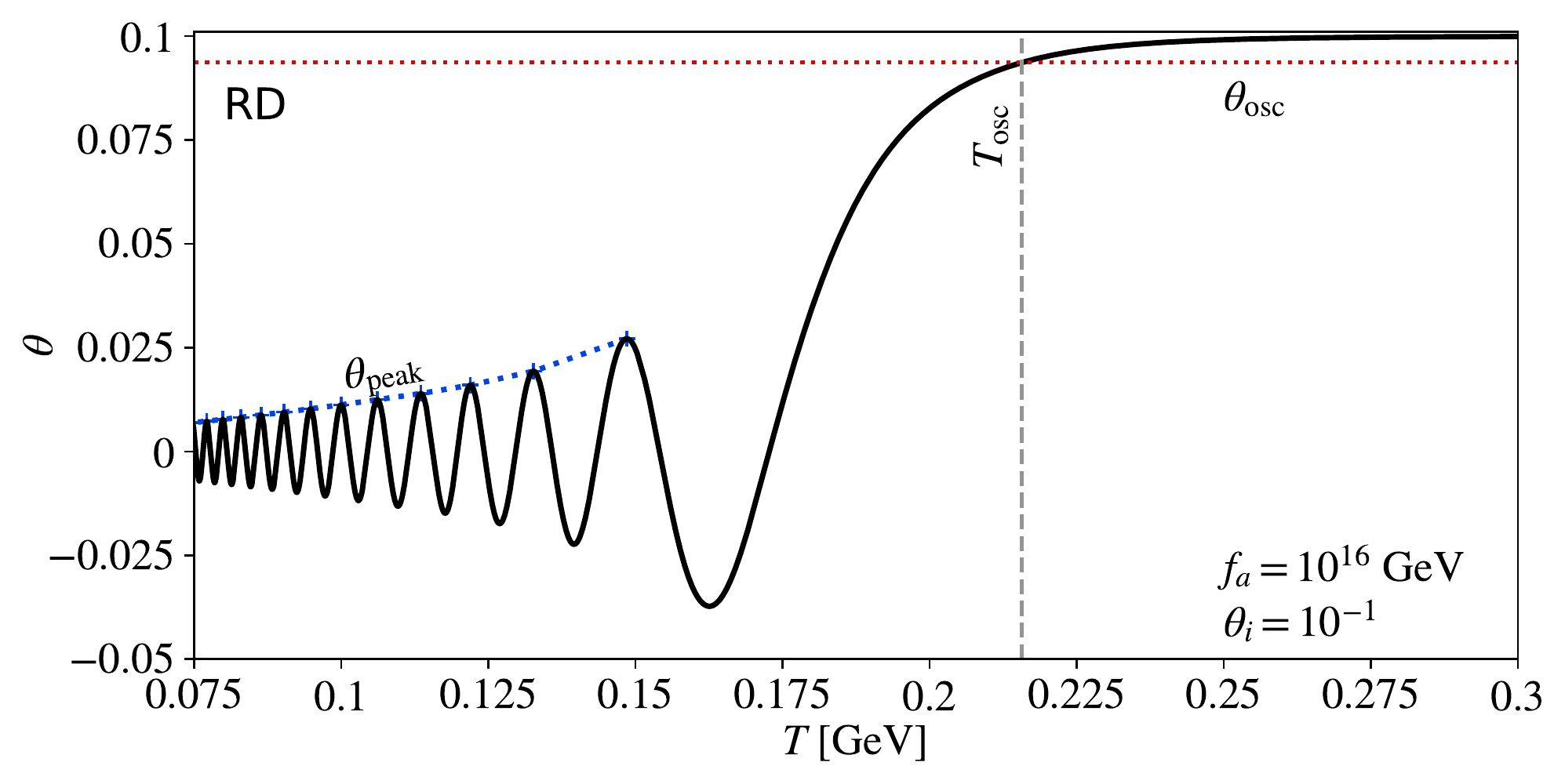}
		\caption{}
		\label{fig:theta_evolution-RD}
	\end{subfigure}
	\caption{The evolution of the axion angle, $\theta$, with the temperature in early matter dominated (a) and radiation dominated (b) cases. For the EMD case, the $\thetai$ is chosen so that $\Omega h^2 = 0.12$.}
	\label{fig:results}
\end{figure}
\begin{figure}[h]
	\begin{subfigure}[]{0.5\textwidth}
		\includegraphics[width=1\textwidth]{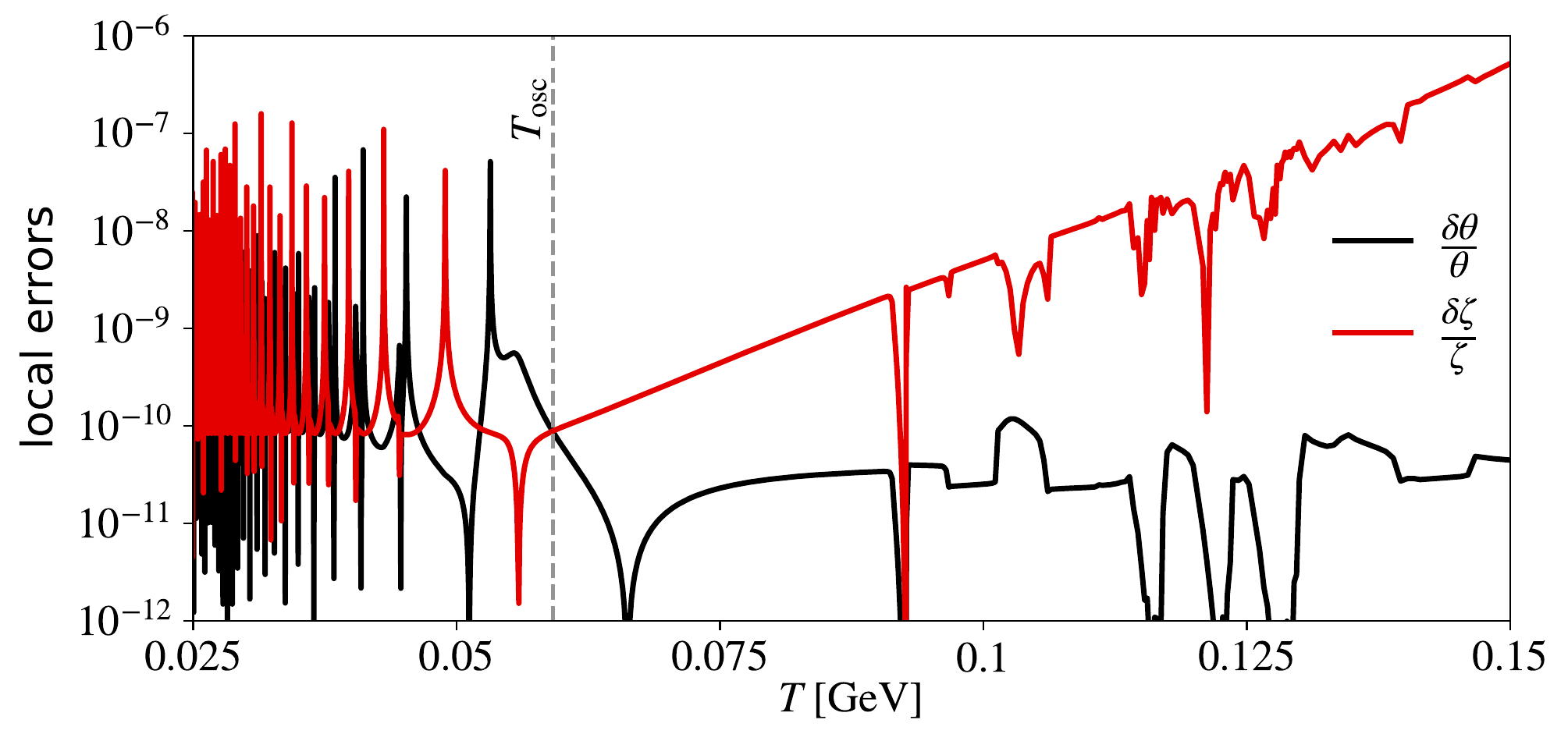}
		\caption{}
		\label{fig:local_errors-EMD}
	\end{subfigure}
	\begin{subfigure}[]{0.5\textwidth}
		\includegraphics[width=1\textwidth]{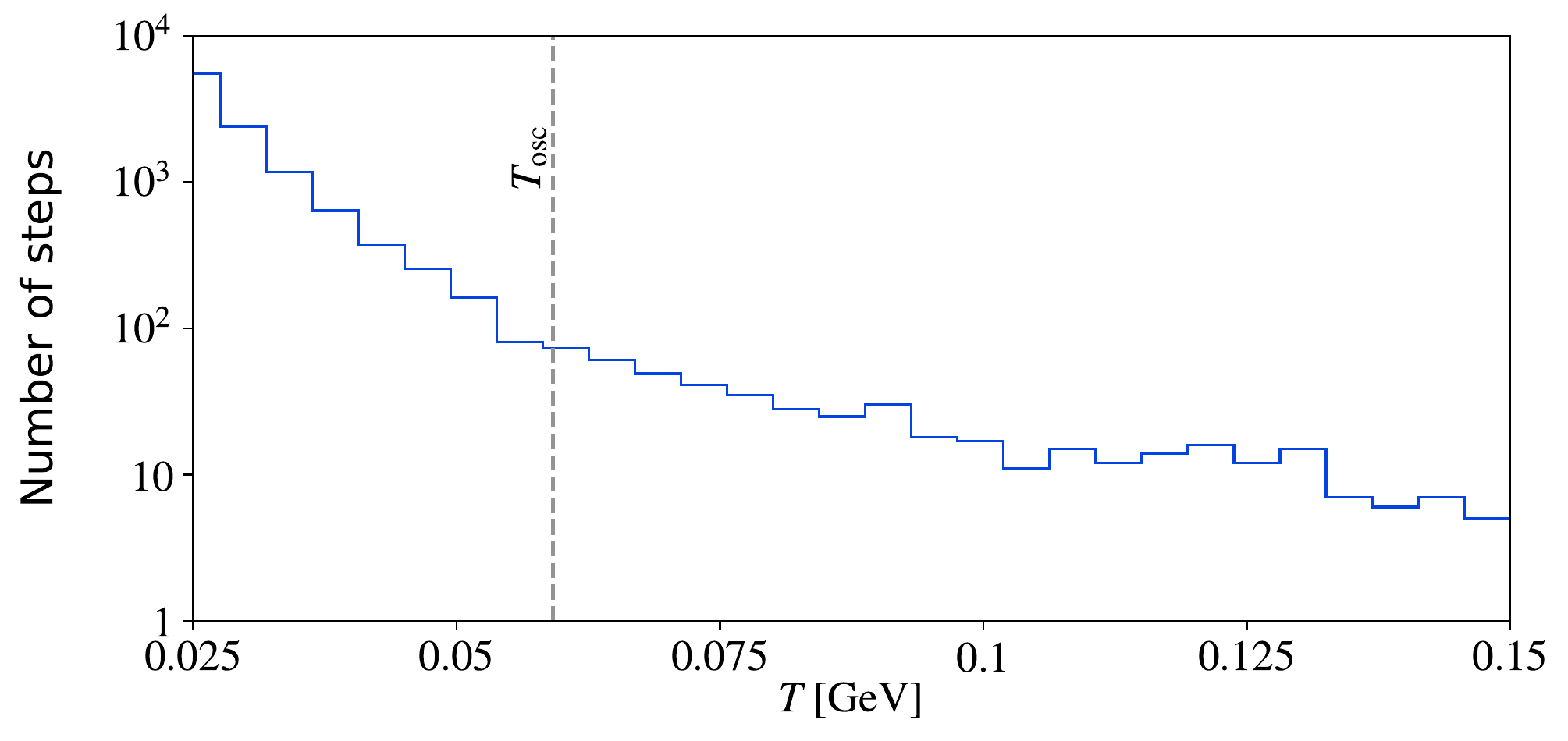}
		\caption{}
		\label{fig:histogram-EMD}
	\end{subfigure}
	\caption{(a) Local relative errors of $\theta$ and $\zeta$ for the temperature range $0.025\GeV \leq T \leq 0.15\GeV$. 
			 (b) Histogram of number of points in the same temperature range.}
	\label{fig:RK_response}
\end{figure}

In \Figs{fig:theta_evolution-EMD} we show the evolution  of $\theta$ for temperatures $T \in [0.025,0.15]~\GeV$, where the vertical line indicates $\Tosc$, while the horizontal one the corresponding value of $\theta$, $\thetaosc$. Also, the blue curve connects the peaks of the oscillation. For comparison, in \Figs{fig:theta_evolution-RD} we also show the evolution of the angle in a radiation dominated Universe with constant entropy (\ie standard cosmological scenario). From these figures, we can see that the effect of an early matter domination reduces the amplitude of oscillation -- due to the injection of entropy -- as well the oscillation temperature -- due to the increase of the Hubble parameter. Furthermore,  the angle at $T=\Tosc$ in the EMD scenario is much smaller that the corresponding value in the standard cosmological case, since the entropy injection causes the scale factor to be larger compared to the scale factor at the same temperature.
Moreover, in \Figs{fig:RK_response}, we show the relative local error of integration as well as a histogram of the number of integration steps for $0.025~\GeV < T < 0.15 ~\GeV$.
The local relative errors, defined in Appendix~\ref{app:RK}, are shown in \Figs{fig:local_errors-EMD}. The black and red lines correspond to $\delta \theta/\theta$ and $\delta \zeta/\zeta$, respectively. This figure indicates that the local errors are relatively well behaved for $T>\Tosc$, and they only start to oscillate violently once the oscillations start. However, the adaptation of the integration step seems to work, as the errors are kept below $\sim 10^{-7}$ (the initial relative error of $\zeta$ is large because $\zeta \approx 0$ before the $\Tosc$).  In order to examine how the step-size is adapted to the difficulty of the problem, we also show a histogram which shows how many integration steps are taken for fixed temperature intervals.~\footnote{More precisely, the histogram is made by dividing the temperature range $0.025~\GeV\lesssim T\lesssim 0.15~\GeV$ to $30$ bins of equal size.} 
In this figure, we see that the number of integration steps increases rapidly for temperatures below the oscillation temperature. 
This is expected, since integration becomes more difficult as the frequency of the oscillation increases -- $\maT/H$ increases rapidly for $T<\Tosc$. That is, the local integration error tends to increase. Thus, in order to reduce the local error, the embedded RK method we employ, reduces the step-size. This means that for $T \lesssim \Tosc$ the number of integration steps increase drastically.
This is the general picture of how adaptation happens. However, one has to experiment with all the available parameters, in order to solve the axion EOM as accurately and fast as possible.  

\section{Acknowledgements}
The author acknowledges support by the Lancaster–Manchester–Sheffield Consortium for Fundamental Physics, under STFC research grant ST/T001038/1.

\section{Summary}
We have introduced \mimes; a header-only library written \CPP that is used to compute the axion (or ALP) relic abundance, by solving the corresponding EOM, in a user defined underlying cosmology. \mimes makes only a few assumptions, which allows the user to explore a wide range ALP and cosmological scenarios. 

In this manuscript, we have provided a detailed explanation on how to use \mimes, by showing examples in both \CPP and \PY (paragraphs~\refs{sec:First_examples,sec:complete_examples}). We have described the user input that is expected, and the various options available (sections~\refs{sec:first_steps,sec:assumptions}). Moreover, in the Appendix, we provide a detailed review of all the internal components that comprise \mimes. We briefly discuss the Runge-Kutta methods that \mimes uses, and show how the user can implement their own. We explain the functionality of all the classes, modules, and utilities. Also, we provide a detailed input and option guide.

In the future, \mimes will be extended in several ways. First, we should implement new functionality that will allow the user to automatically compare against various experimental data (although, there is already an available module~\cite{2020-AxionLimits} that can be used for this goal). This is going to be helpful, as the user will only need to use one program (or script) to compute what is needed. We also aim to supplement \mimes with the option to produce ALPs via interactions of the plasma (\eg freeze-out/in), which may be useful in certain cases.  Moreover, a later version of \mimes may allow the user to define different initial condition for $\zeta$ as well, since there are cases where this is needed (\eg~\cite{Co:2019jts}). Finally, \mimes will continue to improve by correcting mistakes, or implementing suggestions by the community.

\setcounter{section}{0}
\section*{Appendix}
\appendix

\renewcommand{\theequation}{\Alph{section}.\arabic{equation}}
\setcounter{equation}{0}  

\section{Basics of embedded Runge-Kutta Mehtods}\label{app:RK}
\setcounter{equation}{0}
Runge-Kutta (RK) methods are employed in order to solve an ordinary differential equation (ODE), or a system of ODEs of first order.~\footnote{Boundary value problems, and higher order differential equations are expressed as first order ODEs, and then solved. Similarly to  \eqs{eq:eom_sys}.}   Although there are some very insightful sources in the literature (\eg~\cite{Hairer,hairer2010solving,10.5555/1403886}) we give a brief overview of them in order to help the user to make appropriate decisions when using \mimes.

The general form of a system of first order of ODEs is
\begin{equation}
	\dfrac{d\vec{y}}{dt}=\vec{f}(\vec{y},t) \;,
	\label{eq:ODE_definition}
\end{equation}
with given initial condition $\vec{y}(0)$. Also, the components of $\vec{y}$ denote the unknown functions. Note that we can always shift $t$ to start at $0$, which simplifies the notation. In order to solve the system of \eqs{eq:ODE_definition}, an RK method uses an iteration of the form 
\begin{equation}
	\vec{y}_{n+1}=\vec{y}_{n}+ h\sum_{i=1}^{s} b_i \ \vec{k}_i \;,
	\label{eq:RK_iter}
\end{equation}
with $n$ denoting the iteration number, $h$ the ``step-size" that is used to progress $t$; $t_{n+1}=t_{n}+h$. Moreover, $s$, $b_i$ and $\vec{k}_i$ define the corresponding RK method. For example,
the classic Euler method is an RK method with $s=1$, $b=1$, and $\vec{k}_1 = \vec{f}(\vec{y_n , t_n})$. Methods with $\vec{k}$ that depends on previous step (\ie $\vec{k} =\vec{k}\lrsb{y_{n},t_{n}}$), are called explicit, while the ones that try to also predict next step (\ie $\vec{k} =\vec{k}\lrsb{y_{n},t_{n}; y_{n+1},t_{n+1}}$) are called implicit.~\footnote{Generally, by substituting  implicit methods $y_{n+1}$ as in \eqs{eq:RK_iter}, we end up with a system of equations that need to be solved in order to compute $\vec{k}$.} 

\subsection{Embedded RK methods}
A large category of RK methods are  the so-called embedded RK methods. These methods make two estimates for the same step simultaneously -- without evaluating  $\vec{k}$ many times within the same iteration. Therefore, together with the iteration of \eqs{eq:RK_iter}, a second estimate is given by
\begin{equation}
	\vec{y}_{n+1}^{*}=\vec{y}_{n}+ h\sum_{i=1}^{s} b_i^{*} \vec{k}_i \;,
	\label{eq:RK_Embedded_iter}
\end{equation}
with $b_i^{*}$ is an extra parameter that characterise the ``embedded" method of different order (typically, one order higher that the estimate~\ref{eq:RK_iter}). The local (for the step $t_{n+1}$) error divided by the scale of the solution, then, estimated as
\begin{equation}
	\Delta \equiv  \sqrt{\frac{1}{N}\sum_{d=1}^{N}\lrb{\frac{y_{n+1 ,d}-y^{*}_{n+1,d}}{ \Lambda_d }}^{2}} \;,
	\label{eq:error_estimate}
\end{equation}
where $n$ the iteration number, $N$ the number of ODEs, $d$ the component of $\vec{y}$, and $\Lambda_d$ defined as 
\begin{equation}
	\Lambda_d = {\tt Atol} + \max\lrb{|y_{n+1, d}|,|y^{*}_{n+1 ,d}|} \; {\tt Rtol} \;,
	\label{eq:RK_scale}
\end{equation}
with {\tt Atol} and {\tt Rtol} the absolute and relative tolerances that characterise the desirable accuracy we want to achieve; user defined values, typically {\tt Atol}$=${\tt Rtol}$\ll 1$.  With these definitions, the desirable error is reached when  $\Delta \lesssim 1$. 

\paragraph{Step-control} The definition~\ref{eq:error_estimate}, allows us to adjust the step-size $h$ in such way that $\Delta \approx 1$. That is, we take trial steps, and $h$ is adapted until $\Delta \approx 1$.  A simple adaptive strategy adjusts step-size, using
\begin{eqnarray}
	h \to \beta \; h \;  {\rm max} \lrsb{f_{\rm min}, {\rm min} \lrb{\Delta^{-\frac{1}{p+1}}, f_{\rm max}}} \;,
	\label{eq:step-control}
\end{eqnarray}
with $p$ the order of the RK method ($p+1$ is the order of the embedded one), $\beta$  a bias factor of the adaptive strategy (typically $\beta$ is close but below $1$), used to adjust the tendency of $h$ to be somewhat smaller than what the step-control predicts. Also, $f_{min}$ and $f_{max}$ are the minimum and maximum allowed factors, respectively, that can multiply $\beta \ h$, used in order to avoid large fluctuations that can destabilise the process. All these parameters are chosen by the user, in order to make the step-control process as aggressive or safe as needed. 

\paragraph{Correspondence between \mimes parameters and RK ones}
The various parameters that \mimes are described in section~\ref{sec:input}. The correspondence between them and the RK parameters is given in table~\ref{tab:RK_mimes_params}.
\begin{table}[t!]
	\centering
	\begin{tabular}{|c|c|}
		\mimes & Runge-Kutta \\
		\hline
		{\tt absolute\_tolerance } & {\tt Atol}  \\
		\hline
		{\tt relative\_tolerance } & {\tt Rtol}  \\
		\hline
		{\tt b} & $\beta$  \\
		\hline
		{\tt fac\_min} & $f_{\rm min}$  \\
		\hline
		{\tt fac\_max} & $f_{\rm max}$  \\
		\hline
	\end{tabular}
	\caption{Correspondence between the user \mimes user input and the RK parameters described in the text.}
	\label{tab:RK_mimes_params}
\end{table}

\subsection{Explicit embedded RK methods}
Explicit methods use only the information of the previous step in order to compute $\vec{k}_i$ from  
\begin{equation}
	\vec{k}_{i}=\vec f\lrBiggb{ \vec{y}_{n}+h \lrBigb{ \sum_{j=1}^{i-1}a_{ij}\vec{k}_{j} }, t_{n}+ c_{i} \; h}\;, \quad \forall i=1,2,\dots, s\;,
	\label{eq:explicit_RK_k}
\end{equation}
with $a_{ij}$, $c_i$, together with $b_i$ and $b_i^{*}$, consist the so-called Butcher tableau of the corresponding method. For explicit methods this is usually presented as 
\begin{equation}
	\begin{array}{c|cccccc}
		0      & 0      &   0   &      0& \dots & 0& 0\\
		c_2    & a_{21} &   0   &   0   & \dots & 0& 0\\
		c_3    & a_{31} & a_{32}&      0& \dots & 0& 0\\
		\vdots & \vdots & \vdots& \vdots&\ddots &\ddots& \vdots\\
		c_s    & a_{s1} & a_{s2}& a_{s3}& \dots & a_{s s-1}& 0 \\
		\hline
		p      & b_1    & b_2  & b_3 & \dots & b_{s-1} & b_s \\
		p+1      & b_1^{\star}    & b_2^{\star}  & b_3^{\star} & \dots & b_{s-1}^{\star} & b_s^{\star}
	\end{array}
	\label{eq:Butcher}
\end{equation}
It should be noted that $c_i = \displaystyle\sum_{j=1}^{s} a_{ij} $.

\subsection{Rosenbrock methods}
Explicit RK methods, encounter instabilities when a system is ``stiff"~\footnote{A definition of stiffness can be found in~\cite{hairer2010solving,Hairer}.}; \eg when it oscillates rapidly, or has different elements at different scales. These problems are somewhat resolved by trying to predict the next step inside $\vec{k}$; \ie in implicit methods. However, then one has to solve a non-linear set of equations in order to compute $\vec{y}_{n+1}$, which is generally a slow process, as the Newton method (or some variation) needs to be applied. However, there exist another way, a compromise between explicit and implicit methods.  Linearly implicit RK methods, usually called Rosenbrock methods -- with popular improvements as Rosenbrock-Wanner methods, introduce  parameters in the diagonal of the Butcher tableau~\ref{eq:Butcher} and linearise the system of  non-linear equations (for details, see~\cite{hairer2010solving}). In these methods, $\vec{k}$ is determined by  
\begin{equation}
	\left(\hat I - \gamma h \hat{J}\right)\cdot \vec{k}_{i}=
	h \vec{f}\Big(\vec{y}_n+\sum_{j=1}^{i-1}a_{ij}\vec{k}_{j},t_n + c_{i}h \Big)+
	h^2 \left(\gamma + \sum_{j=1}^{i-1}\gamma_{ij}\right)\dfrac{\partial \vec{f} }{\partial t}+
	h \hat J \cdot \sum_{j=1}^{i-1}\gamma_{ij} \vec{k}_{j}\; ,
	\label{eq:Ros_k}
\end{equation}
which is written in such a way that everything is evaluated at $t = t_{n}$. In \eqs{eq:Ros_k},  $\hat J = \dfrac{\partial \vec f}{ \partial \vec y}$ the Jacobian of the system of ODEs, $\hat I$ the unit matrix with dimension equal to the number of ODEs. Moreover, $\gamma$ and $\gamma_{ij}$ are parameters that characterise the method (along with $a_{ij}$, $c_i$,  $b_i$, and $b_i^{*}$).

\paragraph{Implementing a new Butcher tableau in {\tt NaBBODES}} As already mentioned, \mimes uses {\tt NaBBODES}, which supports the implementation of new Butcher tableaux. This is done by adding a new class (or struct) inside the header file {\tt METHOD.hpp} that can be found in {\tt \mimes/src/NaBBODES/RKF} for the explicit RK and  {\tt \mimes/src/NaBBODES/Rosenbrock} for the Rosenbrock embedded methods. All the new parameters must be {\tt public},  {\tt constexpr static} variables, of a type that is the template parameter of the method. For example, the Heun-Euler method can be implemented by adding the following code in  {\tt \mimes/src/NaBBODES/RKF/METHOD.hpp}
\begin{cpp}
	/*-----Implementation of the Heun-Euler method-----*/
	/*-----It shouldn't be used for MiMeS, since it is likely to fail-----*/
	
	//LD is the numeric type (\ie double, or long double).
	template<class LD>
	//use struct because its variables are public by default.
	struct HeunEuler{  
		static constexpr unsigned int s=2; // HeunEuler is a 2-stage RK
		static constexpr unsigned int p=1; // first order, with embedded ?$p+1=2$?

		//these are aliases for the arrays
		using arr=std::array<LD,s>;
		using arr2=std::array<std::array<LD,s>,s>;
		
		static constexpr arr b={0.5,0.5}; // this is ?$b_i$?
		static constexpr arr bstar={0.5,0.5}; // this is ?$b_i^{*}$?
		static constexpr arr c={0,1}; // remember that ?$c_i = \displaystyle\sum_{j=1}^{s} a_{ij}$?
		
		// this is ?$a_{ij}$?
		static constexpr arr2 a={
			arr{0,0},
			arr{1,0}
		};			
	};
\end{cpp}

In order to implement the ROS3w~\cite{RangAngermann2005} method, one can add the following code in the header file {\tt \mimes/src/NaBBODES/Rosenbrock/METHOD.hpp}
\begin{cpp}
	/*-----Implementation of the ROS3w method-----*/
	/*-----It shouldn't be used, since it is likely to fail to produce a result-----*/
	
	
	//LD is the numeric type (?\ie? double, or long double).
	template<class LD>
	struct ROS3w{
		static constexpr unsigned int s=3; // 3-stage
		static constexpr unsigned int p=2; // second order (embedded order is ?$p+1$?)
		
		//aliases for the arrays
		using arr=std::array<LD,s>;
		using arr2=std::array<std::array<LD,s>,s>;

		static constexpr arr b={0.25,0.25,0.5 };  // this is ?$b_i$?
		// this is ?$b_i^{*}$?
		static constexpr arr bstar={ 0.746704703274,0.1144064078371,0.1388888888888};  
		
		static constexpr arr c={0,2/3.,4/3.}; // remember that ?$c_i = \displaystyle\sum_{j=1}^{s} a_{ij}$?
		static constexpr LD gamma=0.4358665215084; // this is ?$\gamma$?
		
		// this is ?$a_{ij}$?
		static constexpr arr2 a={
			arr{0,0,0},
			arr{2/3.,0,0},
			arr{2/3.,2/3.,0}	
		};
	
		// this is ?$\gamma_{ij}$?	
		static constexpr arr2 g={
			arr{0,0,0},
			arr{0.3635068368900,0,0},
			arr{-0.8996866791992,-0.1537997822626,0}
		};
	};
\end{cpp}

\paragraph{How to compile \mimes in order to use the newly implemented method} Once a new Butcher tableau is implemented, the \cppin{mimes::Axion} class can use it. This class, just needs the name of method assigned to the corresponding template argument; {\tt Method}. A convenient way to do this, is to define a macro using the {\tt -D} flag of the compiler, and use macro as the corresponding template parameter of the \cppin{mimes::Axion} class. Alternatively, if one uses the {\tt makefile} files, a method can be chosen by adding it in the corresponding {\tt Definitions.mk} file; \eg as {\tt METHOD=ROS3w} for the {\tt ROW3} method.~\footnote{One needs to make sure to use the correct \cppin{Solver} template argument (or \cppin{SOLVER} variable in the {\tt Definitions.mk} files), otherwise compilation will fail.}

\section{\CPP classes}\label{app:classes}
\setcounter{equation}{0}
\mimes is designed as an object-oriented header-only library. That is, all the basic components of the library are defined as classes inside header files. All the classes relevant to the use of \mimes are under the namespace {\tt mimes}.

\subsection{{\tt Cosmo} class}
The \cppin{mimes::Cosmo<LD>} class is responsible for interpolation of the various quantities of the plasma. Its header file is {\tt \mimes.src/Cosmo/Cosmo.hpp}, and needs to be included in order to use this class. The template parameter \cppin{LD} is the numeric type that will be used, \eg \cppin{double}. The constructor of this class is
\begin{cpp}
	template<class LD>
	mimes::Cosmo<LD>(std::string cosmo_PATH, LD minT=0, LD maxT=mimes::Cosmo<LD>::mP)
\end{cpp}
The argument {\tt cosmo\_PATH} is the path of the data file that contains $T$ (in $\GeV$), $\heff$, $\geff$, with increasing $T$. The parameters {\tt minT} and {\tt maxT} are minimum and maximum interpolation temperatures. These temperatures are just limits, and the actual interpolation is done between the closest temperatures in the data file. Moreover, beyond the interpolation temperatures, both $\heff$ and $\geff$ are assumed to be constants.   

Interpolation of the RDOF, allows us to define various quantities related to the plasma; \eg the entropy density is defined as $s = \dfrac{2\pi^2}{45} \heff T^3$. These quantities are given as the member functions:
\begin{itemize}
	\item \cppin{ template<class LD> LD mimes::Cosmo<LD>::heff(LD T)}: $\heff$ as a function of $T$.
	\item \cppin{ template<class LD> LD mimes::Cosmo<LD>::geff(LD T)}: $\geff$ as a function of $T$.
	\item \cppin{ template<class LD> LD mimes::Cosmo<LD>::dheffdT(LD T)}: $d\heff/dT$ as a function of $T$.
	\item \cppin{ template<class LD> LD mimes::Cosmo<LD>::dgeffdT(LD T)}: $d\geff/dT$ as a function of $T$.
	\item \cppin{ template<class LD> LD mimes::Cosmo<LD>::dh(LD T)}: $\delta_h = 1 + \frac{1}{3} \frac{d\log \heff}{d\log T}$ as a function of $T$.
	\item \cppin{ template<class LD> LD mimes::Cosmo<LD>::s(LD T)}: The entropy density of the plasma as a function of $T$.
	\item \cppin{ template<class LD> LD mimes::Cosmo<LD>::rhoR(LD T)}: The energy density of the plasma as a function of $T$.
	\item \cppin{ template<class LD> LD mimes::Cosmo<LD>::Hubble(LD T)}: The Hubble parameter assuming radiation dominated expansion as a function of $T$.
\end{itemize}

Moreover, there are several cosmological quantities are given as members variables:
\begin{itemize}
	\item \cppin{template<class LD> constexpr static LD mimes::Cosmo<LD>::T0}: CMB temperature today~\cite{Zyla:2020zbs} in $\GeV$.
	\item \cppin{ template<class LD> constexpr static LD mimes::Cosmo<LD>::h_hub}: Dimensionless Hubble constant~\cite{Zyla:2020zbs}.
	\item \cppin{ template<class LD> constexpr static LD mimes::Cosmo<LD>::rho_crit}: Critical density~\cite{Zyla:2020zbs} in $\GeV^3$.
	\item \cppin{ template<class LD> constexpr static LD mimes::Cosmo<LD>::relicDM_obs}: Central value of the measured DM relic abundance~\cite{Planck:2018vyg}.
	\item \cppin{ template<class LD> constexpr static LD mimes::Cosmo<LD>::mP}: Planck mass~\cite{Zyla:2020zbs} in $\GeV$.
\end{itemize}

\subsection{{\tt AnharmonicFactor} class}
The class \cppin{mimes::AnharmonicFactor<LD>} is responsible interpolating the anharmonic factor as defined in \eqs{eq:anharmonic_f}. The corresponding header file is {\tt \mimes/src/AnharmonicFactor/AnharmonicFactor.hpp}. 

The constructor of this class is
\begin{cpp}
	template<class LD>
	mimes::AnharmonicFactor<LD>(std::string anharmonic_PATH)
\end{cpp}
Again, the template argument {\tt LD} is a numeric type, and the {\tt anharmonic\_PATH} string is the path of the data file with data for $\thetamax$ (which should be in increasing order) and $f(\thetamax)$.

The member function that \mimes uses is the overloaded call operator
\begin{cpp}
	template<class LD> LD mimes::AnharmonicFactor<LD>::operator()(LD theta_peak)
\end{cpp}
This function returns the value of the anharmonic factor at $\thetamax=${\tt theta\_peak}. Although, there is no need to call this function beyond the interpolation limits (as long as the data file contains $0\leq \thetamax \leq \pi$), it is important to note that the anharmonic factor is taken to be constant beyond these limits.

\subsection{{\tt AxionMass} class}
The \cppin{mimes::AxionMass<LD>} class is responsible for the definition of the axion mass. The header file of this class is {\tt \mimes/src/AxionMass/AxionMass.hpp}. Its usage and member functions are described in the examples given in sections~\refs{sec:First_examples,sec:complete_examples}. However, it would be helpful to outline them here.

The class has two constructors. The first one is
\begin{cpp}
	template<class LD>
	mimes::AxionMass<LD>(std::string chi_PATH, LD minT=0, LD maxT=mimes::Cosmo::mP)
\end{cpp}
The first argument, {\tt chi\_PATH}, is the path to a data file that contains two columns; $T$ (in $\GeV$) and $\chi$ (in $\GeV^4$), with increasing $T$. The arguments {\tt minT} and {\tt maxT} are the interpolation limits. These limits are used in order to stop the interpolation in the closest temperatures that exist in the data file. That is the actual interpolation limits are $T_{\min}\geq${\tt minT} and $T_{\max}\leq${\tt maxT}. Beyond these limits, by default, the axion mass is assumed to be constant. However, this can be changed by using the member functions
\begin{cpp}
	void set_ma2_MIN(std::function<LD(LD,LD)> ma2_MIN)
	void set_ma2_MAX(std::function<LD(LD,LD)> ma2_MAX)
\end{cpp}
Here, {\tt ma2\_MIN} and {\tt ma2\_MAX} are functors that define the axion mass squared beyond the interpolation limits. In order to ensure that the axion mass is continuous, usually we need $T_{\min}$, $T_{\max}$, $\chi(T_{\rm min})$, and $\chi(T_{\rm max})$. These values can be obtained using the member functions
\begin{itemize}
	\item \cppin{template<class LD> LD mimes::AxionMass<LD>::getTMin()}: This function returns the minimum interpolation temperature, $T_{\rm min}$. 
	\item \cppin{template<class LD> LD mimes::AxionMass<LD>::getTMax()}: This function returns the maximum interpolation temperature, $T_{\rm max}$.
	\item \cppin{template<class LD> LD mimes::AxionMass<LD>::getChiMin()}: This function returns $\chi(T_{\rm min})$.
	\item \cppin{template<class LD> LD mimes::AxionMass<LD>::getChiMax()}: This function returns $\chi(T_{\rm max})$.
\end{itemize}

An alternative way to define the axion mass is via the constructor
\begin{cpp}
	template<class LD>
	mimes::AxionMass<LD>(std::function<LD(LD,LD)> ma2)
\end{cpp}
Here, the only argument is the axion mass squared, $\maT$, defined as a callable object.

Once an instance of the class is defined, we can get $\maT^2$ using the member function
\begin{cpp}
	template<class LD>	LD mimes::AxionMass<LD>::ma2(LD T, LD fa)
\end{cpp}
We should note that {\tt ma2} is a public \cppin{std::function<LD(LD,LD)>} member variable. Therefore, it can be assigned using the assignment operator. However,  in order to change its definition, we can also use the following member function:
\begin{cpp}
	template<class LD> void mimes::AxionMass<LD>::set_ma2(std::function<LD(LD,LD)> ma2)
\end{cpp}

\subsection{{\tt AxionEOM} class}
The \cppin{mimes::AxionEOM<LD>} class is not useful for the user. However, it is responsible for the interpolation of the underlying cosmology, and the definition of the axion EOM~\ref{eq:eom_sys}, which is passed to the ODE solver of {\tt NaBBODES}. 

The constructor of the class is
\begin{cpp}
	template<class LD>
	mimes::AxionEOM<LD>(LD fa, LD ratio_ini, std::string inputFile, AxionMass<LD> *axionMass)
\end{cpp}
The role of the arguments are discussed in section~\refs{sec:cpp_first_example,sec:run_time_input} as well as in table~\ref{tab:AxionSolve-input}. One the instance is created, the interpolations are constructed by calling the member function
\begin{cpp}
	template<class LD> void mimes::AxionEOM<LD>::makeInt()
\end{cpp}
Then, the temperature as a function of $u=\log a/\ai$, is given via the member function 
\begin{cpp}
	template<class LD> LD mimes::AxionEOM<LD>::Temperature(LD u)
\end{cpp}
Another useful member function is  
\begin{cpp}
	template<class LD> LD mimes::AxionEOM<LD>::logH2(LD u)
\end{cpp}
This function returns $\log H^2$  as a function of $u$. Moreover, its derivative, $\frac{d \log H^2}{du}$, is computed using 
\begin{cpp}
	template<class LD> LD mimes::AxionEOM<LD>::dlogH2du(LD u)
\end{cpp}
It should be noted that the highest interpolation temperature is determined by {\tt ratio\_ini} while the lower interpolation temperature is the one given in the data file {\tt inputFile}. Beyond these limits, all functions are assumed to be constant. Therefore, one should be careful, and choose an appropriate {\tt ratio\_ini}, and provide a lower temperature at which any entropy injection has stopped and the axion has reached its adiabatic evolution.

Finally, the actual EOM is given an overloaded call operator
\begin{cpp}
	template<class LD> 
	void mimes::AxionMass<LD>::operator()(std::array<LD,2> &lhs, std::array<LD,2> &y, LD u)
\end{cpp}
Here, the inputs are $u=\log a/\ai$, {\tt y[0]}$=\theta$, and {\tt y[1]}$=\zeta$; which are used to calculate the components of the EOM, with {\tt lhs[0]}$=\frac{d \theta}{d u}$ and {\tt lhs[1]}$=\frac{d \zeta}{d u}$.

\subsection{{\tt Axion} class}
The \cppin{mimes::Axion<LD,Solver,Method>} class is the class that combines all the others, and actually solves the axion EOM~\ref{eq:eom_sys}. Its header file is {\tt \mimes/src/Axion/AxionSolve.hpp}, and its constructor is
\begin{cpp}
 	template<class LD, const int Solver, class Method>
	mimes::Axion< LD, Solver, Method >(LD theta_i, LD fa, LD umax, LD TSTOP, 
			LD ratio_ini, unsigned int N_convergence_max, LD convergence_lim, 
			std::string inputFile, AxionMass<LD> *axionMass, LD initial_step_size=1e-2, 
			LD minimum_step_size=1e-8, LD maximum_step_size=1e-2, LD absolute_tolerance=1e-8, 
			LD relative_tolerance=1e-8, LD beta=0.9, LD fac_max=1.2, LD fac_min=0.8, 
			unsigned int maximum_No_steps=10000000)
\end{cpp}
The various arguments are discussed in section~\refs{sec:cpp_first_example,sec:run_time_input}; and outlined in table~\ref{tab:AxionSolve-input}.

The member function responsible for solving the EOM is
\begin{cpp}
	template<class LD, const int Solver, class Method>
	void mimes::Axion< LD, Solver, Method>::solveAxion()
\end{cpp}
Once this function finishes, the results are stored in several member variables.

The quantities $a/\ai, \ T, \ \theta, \ \zeta, \rho_a$, at the integration  steps are stored in
\begin{cpp}
	template<class LD, const int Solver, class Method> 
	std::vector< std::vector<LD> > mimes::Axion< LD, Solver, Method>::points
\end{cpp}

The quantities $a/\ai, \ T, \ \theta, \ \zeta, \rho_a, \ J$, at the peaks of the oscillation are stored in 
\begin{cpp}
	template<class LD, const int Solver, class Method> 
	std::vector< std::vector<LD> > mimes::Axion< LD, Solver, Method>::peaks
\end{cpp}
Note that these points are computed using linear interpolation between two integration points with a change in the sign of $\zeta$.

The local integration errors for $\theta$ and $\zeta$ are stored in
\begin{cpp}
	template<class LD, const int Solver, class Method> 
	std::vector<LD> mimes::Axion< LD, Solver, Method>::dtheta
	
	template<class LD, const int Solver, class Method> 
	std::vector<LD> mimes::Axion< LD, Solver, Method>::dzeta
\end{cpp}
Moreover, the oscillation temperature, $\Tosc$, and the corresponding values of $a/\ai$ and $\theta$ are given in
\begin{cpp}
	template<class LD, const int Solver, class Method>
	LD mimes::Axion< LD, Solver, Method>::T_osc
	
	template<class LD, const int Solver, class Method>
	LD mimes::Axion< LD, Solver, Method>::a_osc

	template<class LD, const int Solver, class Method>
	LD mimes::Axion< LD, Solver, Method>::theta_osc
\end{cpp}
Also, the entropy injection between the last peak ($T=T_{\rm peak}$) and today ($T=T_0$), $\gamma$ (defined as in \eqs{eq:entropy_injection_gamma}), is given in 
\begin{cpp}
	template<class LD, const int Solver, class Method>
	LD mimes::Axion< LD, Solver, Method>::gamma
\end{cpp}
The relic abundance is stored in the following member variable 
\begin{cpp}
	template<class LD, const int Solver, class Method>
	LD mimes::Axion< LD, Solver, Method>::relic
\end{cpp}

We can set another initial condition, $\thetai$, using 
\begin{cpp}
	template<class LD, const int Solver, class Method>
	void mimes::Axion< LD, Solver, Method>::setTheta_i(LD theta_i)
\end{cpp}
We should note that running this function all variables are cleared. So we lose all information about the last time {\tt axionSolve()} ran. 

In case the mass of the axion is changed, we also need to remake  the interpolation (\ie run \cppin{mimes::AxionEOM::makeInt()}). This is done using
\begin{cpp}
	template<class LD, const int Solver, class Method>
	void mimes::Axion< LD, Solver, Method>::restart()
\end{cpp}
Again, this function clears all member variables. So it should be used with caution.

Finally, there is static \cppin{mimes::Cosmo<LD>} member variable
\begin{cpp}
	template<class LD, const int Solver, class Method>
	static mimes::Cosmo<LD> mimes::Axion< LD, Solver, Method>::plasma
\end{cpp}
This variable can be used without an instance of the \cppin{mimes::Axion<LD,Solver,Method>} class.

\section{\mimes~  \PY interface}\label{app:modules}
\setcounter{equation}{0}

The various \PY modules, classes, and functions are designed to work exactly in the same way as the ones in \CPP. All the modules are located in \pyin{src/interfacePY}, so it is helpful to add the {\tt \mimes/src} path to the system path at the top of every script that uses \mimes. This is done by adding
\begin{py}
	from sys import path as sysPath
	sysPath.append('path_to_src')
\end{py}

The available models are {\tt Cosmo}, {\tt AxionMass}, and {\tt Axion}, each defines a class with the same name.

\subsection{{\tt Cosmo} class}
The {\tt Cosmo} module defines the \pyin{Cosmo} class, which contains information about the plasma. The relevant shared library ({\tt lib/libCosmo.so}) is obtained by compiling {\tt \mimes/src/Cosmo/Cosmo.cpp} using {\tt make lib/libCosmo.so}.

The class can be imported by running 
\begin{py}
	from interfacePy.Cosmo import Cosmo
\end{py}
Its constructor is
\begin{py}
	Cosmo(cosmo_PATH, minT=0, maxT=mP)
\end{py}
The argument {\tt cosmo\_PATH} is the path (a string) of a data file that contains $T$ (in $\GeV$), $\heff$, $\geff$, with accenting $T$. The second and third arguments, {\tt minT} and {\tt maxT}, are minimum and maximum interpolation temperatures, with the interpolation being between the closest temperatures in the data file. Moreover, beyond these limits, both $\heff$ and $\geff$ are assumed to be constants. It is important to note that the class creates a \cppin{void} pointer that gets recasted to \cppin{mimes::Cosmo<LD>} in order to call the various member functions. This means that once an instance of \pyin{Cosmo} is no longer needed, it must be deleted, in order to free the memory that it occupies. An instance, say {\tt cosmo}, is deleted using
\begin{py}
	del cosmo
\end{py}

The member functions of this class are:
\begin{itemize}
	\item \pyin{Cosmo.heff(T)}: $\heff$ as a function of $T$.
	\item \pyin{Cosmo.geff(T)}: $\geff$ as a function of $T$.
	\item \pyin{Cosmo.dheffdT(T)}: $d\heff/dT$ as a function of $T$.
	\item \pyin{Cosmo.dgeffdT(T)}: $d\geff/dT$ as a function of $T$.
	\item \pyin{Cosmo.dh(T)}: $\delta_h = 1 + \frac{1}{3} \frac{d\log \heff}{d\log T}$ as a function of $T$.
	\item \pyin{Cosmo.s(T)}: The entropy density of the plasma as a function of $T$.
	\item \pyin{Cosmo.rhoR(T)}: The energy density of the plasma as a function of $T$.
	\item \pyin{Cosmo.Hubble(T)}: The Hubble parameter assuming radiation dominated expansion as a function of $T$.
\end{itemize}

The several cosmological quantities are given as members variables:
\begin{itemize}
	\item \pyin{Cosmo.T0}: CMB temperature today~\cite{Zyla:2020zbs} in $\GeV$.
	\item \pyin{Cosmo.h_hub}: Dimensionless Hubble constant~\cite{Zyla:2020zbs}.
	\item \pyin{Cosmo.rho_crit}: Critical density~\cite{Zyla:2020zbs} in $\GeV^3$.
	\item \pyin{Cosmo.relicDM_obs}: Central value of the measured DM relic abundance~\cite{Planck:2018vyg}.
	\item \pyin{Cosmo.mP}: Planck mass~\cite{Zyla:2020zbs} in $\GeV$.
\end{itemize}

Note that these values can be directly imported from the module, without declaring an instance of the class, as 
\begin{py}
	from interfacePy.Cosmo import T0, h_hub, rho_crit, relicDM_obs, mP
\end{py}

\subsection{{\tt AxionMass} class}
The \pyin{AxionMass} class is defined in the module with the same name that can be found in the directory {\tt \mimes/src/interfacePy/AxionMass}. This class is responsible for the definition of the axion mass. This module loads the corresponding shared library from {\tt \mimes/lib/libma.so}, which is created by compiling {\tt \mimes/src/AxionMass/AxionMass.cpp} using \run{make lib/libma.so}. 
Its usage is described in the examples given in sections~\refs{sec:First_examples,sec:complete_examples}. Moreover, this class is used in the same way as \cppin{mimes::AxionMass<LD>}. However, we should append in this section the definition of its member functions in \PY.

The class is imported using 
\begin{py}
	from interfacePy.AxionMass import AxionMass
\end{py}

The constructor is
\begin{py}
	AxionMass(*args)
\end{py}
and can be used in two different ways. 

First, one can pass three arguments, \ie
\begin{py}
	AxionMass(chi_PAT, minT, maxT)
\end{py}
The first argument is the path to a data file that contains two columns; $T$ (in $\GeV$) and $chi$ (in $\GeV^4$), with increasing $T$. The arguments {\tt minT} and {\tt maxT} are the interpolation limits. These limits are used in order to stop the interpolation in the closest temperatures in {\tt chi\_PATH}. That is the actual interpolation limits are $T_{\min}\geq${\tt minT} and $T_{\max}\leq${\tt maxT}. Beyond these limits, by default, the axion mass is assumed to be constant. However, this can be changed by using the member functions
\begin{py}
	AxionMass.set_ma2_MIN(ma2_MIN)
	AxionMass.set_ma2_MAX(ma2_MAX)
\end{py}
Here, {\tt ma2\_MIN(T,fa)} and {\tt ma2\_MAX(T,fa)}, are functions (not any callable object), should take as arguments {\tt T} and {\tt fa}, and return the axion mass squared beyond the interpolation limits. In order to ensure that the axion mass is continuous, usually we need $T_{\min}$, $T_{\max}$, $\chi(T_{\rm min})$, and $\chi(T_{\rm max})$. These values can be obtained using the member functions
\begin{itemize}
	\item \pyin{AxionMass.getTMin()}: This function returns the minimum interpolation temperature, $T_{\rm min}$. 
	\item \pyin{AxionMass.getTMax()}: This function returns the maximum interpolation temperature, $T_{\rm max}$.
	\item \pyin{AxionMass.getChiMin()}: This function returns $\chi(T_{\rm min})$.
	\item \pyin{AxionMass.getChiMax()}: This function returns $\chi(T_{\rm max})$.
\end{itemize}

An alternative way to define the axion mass is via the constructor
\begin{cpp}
	AxionMass(ma2)
\end{cpp}
Here, {\tt ma2(T,fa)} is a function (not any callable object) that takes $T$ (in $\GeV$) and $f_a$, and returns  $m_a^2(T)$ (in $\GeV$). As in the other \PY classes, once the instances of this class are no longer needed, they must be deleted using the destructor, \pyin{del}.

The member function that returns $\maT^2$ is
\begin{py}
	AxionMassma2(T,fa)
\end{py}
We should note that another {\tt ma2} can be changed using the following member function:
\begin{py}
	AxionMass.set_ma2(ma2)
\end{py}
Again, {\tt ma2(T,fa)} is a function (not any callable object) that takes $T$ (in $\GeV$) and $f_a$, and returns  $m_a^2(T)$ (in $\GeV$).

\subsection{{\tt Axion} class}
The class \pyin{Axion}, solves the axion EOM~\ref{eq:eom_sys}. This class is defined in {\tt \mimes/interfacePy/Axion/Axion.py}, and imports the corresponding shared library. This library is compiled by running {\tt make lib/Axion\_py.so}, and its source file is {\tt \mimes/src/Axion/Axion-py.cpp}. As in the previous classes, its usage is similar to the \CPP version.

Its constructor is 
\begin{py}
	Axion(theta_i, fa, umax, TSTOP, 
			ratio_ini, unsigned int N_convergence_max, convergence_lim, 
			inputFile, axionMass, initial_step_size=1e-2, 
			minimum_step_size=1e-8, maximum_step_size=1e-2, absolute_tolerance=1e-8, 
			relative_tolerance=1e-8, beta=0.9, fac_max=1.2, fac_min=0.8, 
			unsigned int maximum_No_steps=10000000)
\end{py}
Again, the various arguments are discussed in section~\refs{sec:cpp_first_example,sec:run_time_input}; and can also be found in table~\ref{tab:AxionSolve-input}. Notice one important difference between this and the \CPP verison of this class; the instance of \pyin{AxionMass}, {\tt axionMass}, is passed by value (there are no pointers in \PY). However, internally, the constructor converts this instance to a pointer, which is then passed to the underlying {\tt C} function responsible for creating the relevant instance.

The member function responsible for solving the EOM is
\begin{py}
	Axion.solveAxion()
\end{py}
Once this function is finished, the following member funcions are available
\begin{itemize}
	\item \pyin{Axion.T_osc}: the oscillation temperature, $\Tosc$, in $\GeV$.
	\item \pyin{Axion.a_osc}: $\dfrac{a}{\ai}$ at the oscillation temperature.
	\item \pyin{Axion.theta_osc}: $\thetaosc$, \ie $\theta$ at $\Tosc$.
	\item  \pyin{Axion.gamma}: the entropy injection between the last peak ($T=T_{\rm peak}$) and today ($T=T_0$), $\gamma$ (defined as in \eqs{eq:entropy_injection_gamma}).
	\item \pyin{Axion.relic} The relic abundance of the axion.
\end{itemize}

The evolution of $a/\ai, \ T, \ \theta, \ \zeta, \rho_a$ at the integration steps, is not automatically accessible to user, but they can be made so using
\begin{py}
	Axion.getPoints()
\end{py}
Then, the following member variables are filled
\begin{itemize}
	\item \pyin{Axion.a}: The scale factor over its initial value, $\dfrac{a}{\ai}$.
	\item \pyin{Axion.T}: The temperature in $\GeV$.
	\item \pyin{Axion.theta}: The axion angle, $\theta$.
	\item \pyin{Axion.zeta}: The derivative of $\theta$, $\zeta \equiv \dfrac{d \theta}{d \log (a/\ai)}$.
	\item \pyin{Axion.rho_axion}: The axion energy density in $\GeV^4$.
\end{itemize}

Moreover, the function
\begin{py}
	Axion.getPeaks()
\end{py}
fills the (\pyin{numpy}) arrays \pyin{Axion.a_peak}, \pyin{Axion.T_peak}, \pyin{Axion.theta_peak}, \pyin{Axion.zeta_peak}, \pyin{Axion.rho_axion_peak}, and \pyin{Axion.adiabatic_invariant} with the quantities $a/\ai, \ T, \ \theta, \ \zeta, \rho_a, \ J$, at the peaks of the oscillation. These points are computed using linear interpolation between two integration points with a change in the sign of $\zeta$.

The local integration errors for $\theta$ and $\zeta$ are stored in	\pyin{Axion.dtheta} and \pyin{Axion.dzeta}, after the following function is run
\begin{py}
	Axion.getErrors()
\end{py}

Another initial condition, $\thetai$, can be used without declaring a new instance using
\begin{py}
	Axion.setTheta_i(theta_i)
\end{py}
We should note that running this function all variables are cleared. 

As in the previous \PY classes, once an instance of the \pyin{Axion} class is no longer needed, it needs to be deleted, by calling the destructor, \pyin{del}. 

\paragraph{Important difference between the \CPP version} Since the axion mass is passed by value in the constructor, a change of the  \pyin{AxionMass} instance has no effect on the \pyin{Axion} instance that uses it. Therefore, if the definition of the axion mass changes, one has to declare a new instance of the \pyin{Axion} class. The new instance can be named using the name of the previous one, if the latter is deleted by running its destructor.

\section{Other modules}\label{app:other_modules}
There are several modules available, that may help the user scan to obtain approximate results using the WKB approximation, sca the parameter space, and plot some results. These modules are not
integral to \mimes, and the user is free to ignore them.

\subsection{{\tt WKB} module}
The {\tt WKB} module can be used to calculate the axion relic abundance using the WKB approximation discussed in section~\ref{sec:Physics}. The module can be imported using 
\begin{py}
	from interfacePy import WKB
\end{py}
It contains the definition of a function that returns the relic abundance using the WKB approximation, which is 
\begin{py}
	WKB.relic(Tosc, theta_osc ,ma2, gamma=1, 
	          cosmo=Cosmo(_PATH_+r'/src/data/eos2020.dat',0,Cosmo.mP))
\end{py}
Here, {\tt Tosc} is the oscillation temperate, {\tt ma2(T,fa)} is $\ma^2(T)$ as a function that takes $T$ and $f_a$ as arguments, {\tt cosmo} an instance of the \pyin{Cosmo} class, and {\tt gamma} the entropy injection (as defined in \eqs{eq:WKB_gamma_def}). 

Moreover, there is a function that helps to determine $\Tosc$ and $\gamma$, which is
\begin{py}
	WKB.getPoints(fa, ma2, inputFile, cosmo=Cosmo(_PATH_+r'/src/data/eos2020.dat',0,Cosmo.mP))
\end{py}
The arguments {\tt fa}, {\tt inputFile},$\fa$, and {\tt cosmo} the path to a file that describes the cosmology (described in table~\ref{tab:AxionSolve-input}), $\fa$ (in $GeV$), an instance of the \pyin{Cosmo} class, respectively.  
This function returns $\gamma$ (the entropy injection between {\tt Tosc} and today) and $\Tosc$.

\subsection{The {\tt ScanScript} module}\label{app:ScanScript}
The \PY interface of \mimes has two simple classes that help to scan over $\thetai$ and $\fa$, in parallel. These two classes can be imported from the module \pyin{ScanScript}.  

Tis module is based on the \pyin{bash} script {\tt \mimes/src/util/parallel\_scan.sh}, which automatically performs scans in parallel. This script is used as
\begin{py}
	bash src/util/parallel_scan.sh executable cpus inputFile
\end{py}  
Here, {\tt executable} is the path to an executable, {\tt cpus} the number of instances of the {\tt executable} to launch simultaneously, and {\tt inputFile} a file that contains the arguments the {\tt executable} expects. This file should contain arguments for the {\tt executable} in each line. This script, then, separates all in arguments in the {\tt inputFile} in batches of size {\tt cpus}, and runs one batch at a time.

\subsubsection{The {\tt Scan} class}\label{app:Scan}
The \pyin{Scan} class writes a time-coded file (so it would be unique) with columns that correspond to $\thetai$, $\fa$ ($\GeV$), $\thetaosc$, $\Tosc$ (in $\GeV$), and $\Omega h^2$, for every combination of $\thetai$ and $\fa$ that are passed as input. 

This class is imported using 
\begin{py}
	from interfacePy.ScanScript import Scan
\end{py}

Its constructor is
\begin{py}
	ScanScript.Scan(cpus,table_fa,table_theta_i,umax,TSTOP,ratio_ini,
					N_convergence_max,convergence_lim,inputFile,
					PathToCppExecutable, break_after=5*60,break_time=5,break_command='',
					initial_step_size=1e-2, minimum_step_size=1e-8, maximum_step_size=1e-2, 
					absolute_tolerance=1e-8, relative_tolerance=1e-8,
					beta=0.9, fac_max=1.2, fac_min=0.8, maximum_No_steps=int(1e7))
\end{py}
Some arguments are the same as in the \pyin{Axion} class, and their definition can be found in table~\ref{tab:AxionSolve-input}. The other arguments are
\begin{itemize}
        \item {\tt cpus}: number of points to run simultaneously (number of cpus available). 
		\item {\tt table\_fa}: table of $\fa$ to scan.
		\item {\tt table\_theta\_i}: table of $\thetai$ to scan
        \item {\tt PathToCppExecutable}: path to an executable that takes {\tt theta\_i},  {\tt fa},  {\tt umax},  {\tt TSTOP},  {\tt ratio\_ini}, {\tt N\_convergence\_max}, {\tt convergence\_lim}, and  {\tt inputFile}; and prints {\tt theta\_i}, {\tt fa}, {\tt theta\_osc}, {\tt T\_osc}, and {\tt relic}.
		\item {\tt break\_after}, {\tt break\_time}: take a break after {\tt break\_after} seconds for {\tt break\_time} seconds.
		\item {\tt break\_command} (optional): before it takes a break, run this system command (this may be a script to send the results
		via e-mail, or back them up).
\end{itemize}
The scan runs using the method
\begin{py}
	Scan.run()
\end{py} 
For every value of $\fa$, this method calls another one \pyin{Scan.run_batch()}, which in turn calls the \pyin{bash} script {\tt \mimes/src/util/parallel\_scan.sh} with arguments {\tt PathToCppExecutable}, {\tt cpus}, and a file that contains all the inputs for {\tt PathToCppExecutable} for all values in \pyin{table_theta}.

As the scan runs, it prints the number of batches that have been evaluated, the mean time it takes to evaluate one batch, and an estimate of the remaining time. It should be noted that if the scan exits before it finishes, the next run will continue from the point at which it was stopped even if the inputs have changed. In order to start from the beginning, the user must delete the file {\tt count.\_mimes\_}.

\subsubsection{The {\tt ScanObs} class}\label{app:ScanObs}
The \pyin{ScanObs} class scans for different values of $\fa$ (given as a table) and finds the value of $\thetai$ closer to the observed DM relic abundance. The result file is a time-coded file with columns correspond that to $\thetai$,  $\fa$ (in $\GeV$), $\thetaosc$, $\Tosc$ (in $\GeV$), and $\Omega h^2$. 
It is imported using 
\begin{py}
	from interfacePy.ScanScript import ScanObs
\end{py}
Its constructor is
\begin{py}
	ScanScript.ScanObs(cpus,table_fa,len_theta,umax,TSTOP,ratio_ini,
					N_convergence_max,convergence_lim,inputFile,axionMass,
					PathToCppExecutable, relic_obs,relic_err_up,relic_err_low,
					cosmo=Cosmo(_PATH_+r'/src/data/eos2020.dat',0,Cosmo.mP),
					break_after=5*60,break_time=5,break_command='',
					initial_step_size=1e-2, minimum_step_size=1e-8, maximum_step_size=1e-2, 
					absolute_tolerance=1e-8, relative_tolerance=1e-8,
					beta=0.9, fac_max=1.2, fac_min=0.8, maximum_No_steps=int(1e7)
\end{py}
Here, {\tt len\_theta} is the number of different values of $\theta$ to be used in the search of a value closer to the central value of the observed DM relic abunance. The arguments {\tt relic\_obs}, {\tt relic\_err\_up}, {\tt relic\_err\_low} are the central value of $\Omega h^2$, and its upper and lower error; respectively. All the other arguments have already been defined previously.  

The scan runs using the method
\begin{py}
	ScanObs.run()
\end{py} 
For every value of $\fa$, this method first calculates the relic abundance for $\thetai \ll 1$, and finds the value of $\thetai$ that would result in $\Omega h^2 = ${\tt relic\_obs}; $\thetai^{\rm (approx)}$. This is easy to do, since for $\thetai \ll 1$, the EOM becomes independent of $\thetai$. 

Then, it creates an array, {\tt table\_theta}, of size {\tt len\_theta} that contains values of $\thetai$ between $Min(0.85 \ \thetai^{\rm (approx)},1)$ and  
$Max(1.2 \ \thetai^{\rm (approx)},\pi)$.  Then it calls \pyin{ScanObs.run_batch()}, which works as \pyin{Scan.run_batch()}.

Similarly to the \pyin{Scan} class, if the scan exits before it finishes, the next run will continue from the point at which it was stopped.
In order to start from the beginning, delete the file {\tt count.\_mimes\_}.

\subsection{\tt FT module}
The {\tt FT} class, defined in {\tt FT} module that can be found in {\tt src/interfacePy/FT}, is used to format the ticks when plotting using {\tt matplotlib}~\cite{Hunter:2007}. 
The class can be imported using
\begin{py}
	from interfacePy.FT import FT 
\end{py} 
The constructor of the class is
\begin{py}
	FT(_M_xticks,_M_yticks, _M_xticks_exception,_M_yticks_exception, _m_xticks,_m_yticks,
			xmin,xmax,ymin,ymax,xscale,yscale)
\end{py}
The various arguments are
\begin{itemize}
	\item \pyin{_M_xticks},\pyin{_M_yticks}: A list for  the major ticks in the x and y axes, respectively.
	\item \pyin{_M_xticks},\pyin{_M_yticks}: A list for  the major ticks in the x and y axes, respectively, for which no number is printed.
	\item \pyin{_m_xticks},\pyin{_m_yticks}: A list for  the minor ticks in the x and y axes, respectively.
	\item \pyin{xmin}, \pyin{xmax} (\pyin{ymin},\pyin{ymax}): Minimum and maximum of the x (y) axes, respectively. 
	\item \pyin{xscale}, \pyin{yscale}: The scale of the x and y axes, respectively. The available values are \pyin{"linear"}, \pyin{"log"}, and \pyin{"symlog"}. 
\end{itemize}

Instances of {\tt FT} should be defined after subplots have been defined, because the code seems clearer. Once this is done, we can format the ticks of a subplot, \pyin{sub}, using  
\begin{py}
	FT.format_ticks(plt,sub)
\end{py}
It is important to not that \pyin{plt} is the {\tt matplotlib.pyplot} module that is usually imported as
\begin{py}
	import matplotlib.pyplot as plt
\end{py}

\subsubsection*{Example of {\tt FT}}
Consider as an example the plot of $f(x) = e^{-\frac{1}{2}x^2}$ for $-5 \leq x \leq 5$. In order to do this, we need to run 
\begin{py}
	import numpy as np#you usually need numpy
	
	#---these are for plots---#
	import matplotlib.pyplot as plt
	
	#load the FT module
	from sys import path as sysPath
	sysPath.append('../../src')
	from interfacePy.FT import FT

	fig=plt.figure(figsize=(9,4))
	fig.subplots_adjust(bottom=0.15, left=0.15, top = 0.95, right=0.9,wspace=0.0,hspace=0.0)
	sub = fig.add_subplot(1,1,1)
	
	x=np.linspace(-5,5,500)
	sub.plot(x,np.exp(-1/2. * x**2))
\end{py}

Now we can format the ticks using \pyin{FT}. For the shake of this example, for the x-axis, we choose as major ticks all integrs except $3$ and $-3$, and minor ticks the halves of each major tick. Also, for the y-axis, we choose to only show major ticks for every $0.1$ in the interval of interest. Finally, we set the plot limits as $x \in [-5,5]$ ad $y \in [0,1]$, usng linear scale for both. This can be done as

\begin{py}
    #set major ticks
	_M_xticks=[i for i in range(-5,6)]
	_M_yticks=[i/10. for i in range(0,11)]
	
	#set major ticks that will not have a label
	_M_xticks_exception=[-3,3]
	_M_yticks_exception=[]
	
	_m_xticks=[i+j/2. for i in range(-5,6) for j in [0,1]]
	_m_yticks=[]  
	ft=FT(_M_xticks, _M_yticks, _M_xticks_exception, _M_yticks_exception, _m_xticks, _m_yticks,
		xmin=-5, xmax=5, ymin=0, ymax=1, xscale='linear', yscale='linear')
	
	ft.format_ticks(plt,sub)    
\end{py}
\begin{figure}[t]
	\centering
		\includegraphics[width=0.65\textwidth]{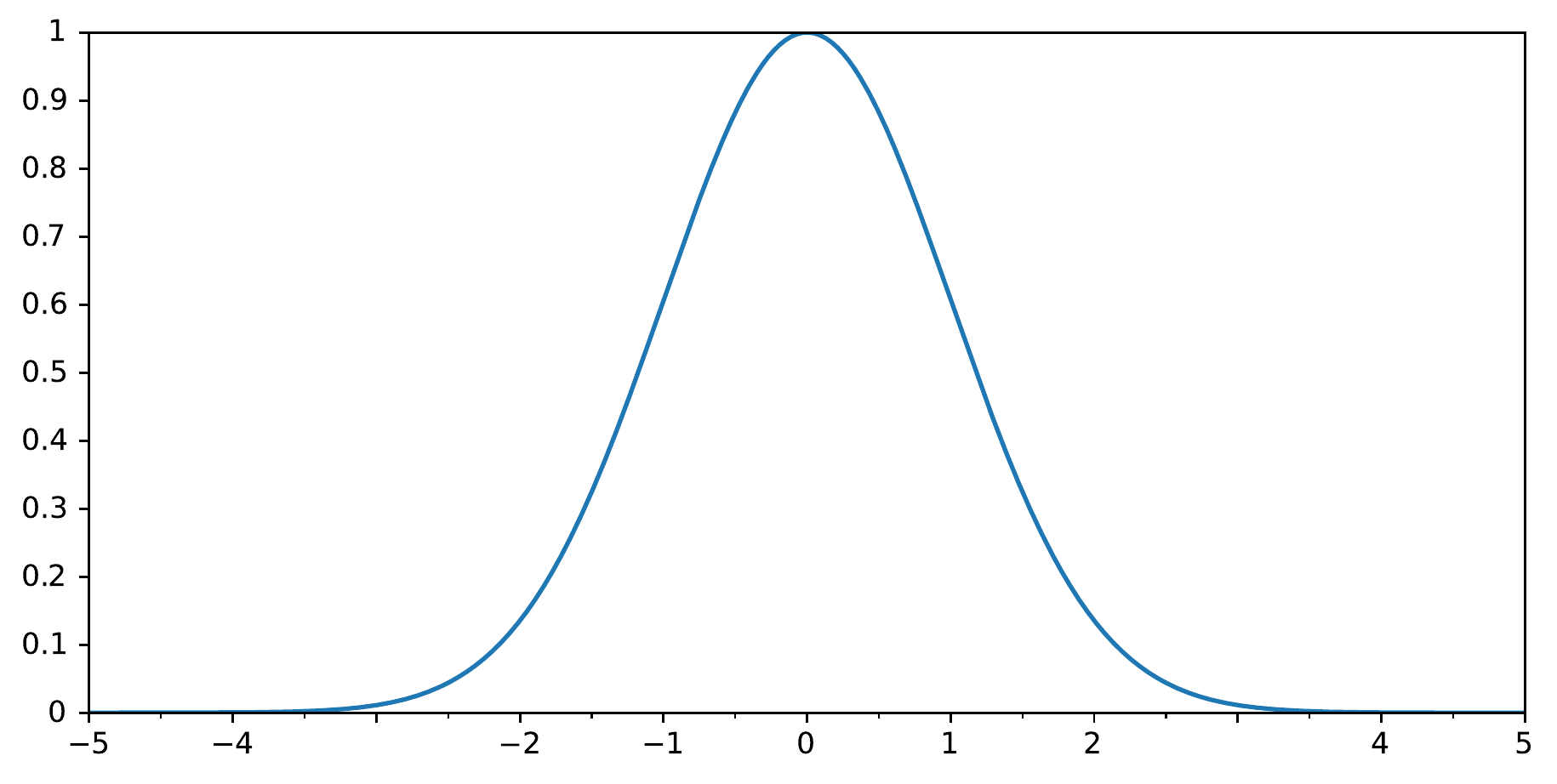}
		\caption{Example plot obtained by using the \pyin{FT} class.}
		\label{fig:FT_example}
\end{figure}
The resulting plot is shown in \Figs{fig:FT_example}.

\section{Utilities}\label{app:util}
\setcounter{equation}{0}
There are various utilities, that can be found {\tt \mimes/src/util}, which can be used to make the use of \mimes easier. In this section we discuss how they work.

\subsection{\tt FormatFile.sh} 
This is a {\tt bash} script that formats a data file so that it is compatible with the interpolation assumptions. It takes a path to a file as an argument, and it returns the same file, sorted (in ascenting order) with respect to the first column, with all duplicate or empty lines as well as the last "new line" (\ie \cppin{\n}) removed. The script is called as 
\begin{cpp}
	bash MiMeS/src/util/FormatFile.sh path_to_dat_file
\end{cpp}
Here {\tt path\_to\_dat\_file} is a path to the  data file we would like to replace with the formatted one. Notice that {\tt FormatFile.sh} must be run in order to ensure that the cosmological input needed by the \cppin{mimes::Axion<LD,Solver,Method>} class, is acceptable.

Moreover, note that {\tt FormatFile.sh} runs every time {\tt configure.sh} is called, in order to format the data files for the RDOFs, $\chi(T)$, and the anharmonic factor, in order to ensure that they comply with what \mimes expects.

\subsection{\tt timeit.sh}
This script takes two arguments. The first should be a path to an executable, the second a file that contains arguments to passed to the executable. The script, then, runs the executable, and prints in \cppin{stderr} the time it took (is seconds) to run it. It is important to note that each argument of the executable should be in a different line of the file.~\footnote{For example , if we have an executable that takes two arguments, let's assume the first is $3$ and the second is "foo", the file should read: \\
\pyin{3}\\
\pyin{foo}
}

The script is called as 
\begin{cpp}
	bash MiMeS/src/util/timeit.sh path_to_executable path_to_dat_file
\end{cpp}
This script is useflul for running the example code in {\tt \mimes/UserSpace/Cpp/Axion/Axion.cpp}, which is compiled by calling \run{make examples} in the root directory of \mimes. The executable that results from the compilation of this source, {\tt \mimes/UserSpace/Cpp/Axion/Axion.run}, expects the same arguments as the constructor of the \cppin{mimes::Axion<LD,Solver,Method>} class (see section~\ref{app:classes}). This means that we can write a file (call it {\tt INPUT}) that contains all these arguments in each line, and simply run  
\begin{cpp}
	bash MiMeS/src/util/timeit.sh MiMeS/UserSpace/Cpp/Axion/Axion.run INPUT
\end{cpp}
The files {\tt KinationInputs}, {\tt MatterInputs}, and {\tt RDInputs}; in {\tt \mimes/UserSpace/Cpp/Axion}, show these files should look like in different cases.

\subsection{{\tt Timer} \CPP class}
The class \cppin{mimes::util::Timer}, defined in {\tt \mimes/src/util/timeit.hpp}, can be used in order to time processes in \CPP. An instance of this class prints  its life-time -- \ie time from the moment it was created until it went out of scope -- in \cppin{stderr}. Instances of this class are intended to be used inside a scope in order to count the time it took to run a block of code. For example the time it takes to compute  $\sum_{i=0}^{10^4} \ i$ can be determined as  
\begin{cpp}
	#include "src/util/timeit.hpp"
	
	int main(){
		auto sum = 0;
		{	
			Timer _timer_;
			for(auto i=0; i<=10000; ++i){ ++sum;}	
		}
	
		return 0;
	}
\end{cpp}

\subsection{{\tt linspace} function}
The function {\tt linspace}, defined in {\tt \mimes/src/util/linspace.hpp}, is used to generate linearly spaced numbers between some boundary. This function is overloaded; there are two functions with the same name and different signature. The first one is 
\begin{cpp}
    /*linspace where the list is pushed in a vector that is pass by reference*/	
	template<class LD>
	void mimes::util::linspace(LD min, LD max, unsigned int length, std::vector<LD> &X)
\end{cpp}
Here, {\tt X} is cleared, and gets filled with  {\tt length} linearly spaced numbers between {\tt min} and {\tt max}.

The overloaded version of this function is 
\begin{cpp}
    /*linspace where the list returned as a vector*/
	template<class LD>
	std::vector<LD> mimes::util::linspace(LD min, LD max, unsigned int length)
\end{cpp}
This function does not get a reference to a \cppin{std::vector<LD>}, instead it return one.

\subsection{{\tt logspace} function}
The function {\tt logspace}, defined in {\tt \mimes/src/util/logspace.hpp}, is used to generate $\log_{10}$-spaced points between a given boundary. The signatures of this function are
\begin{cpp}
    /*logspace where the list is pushed in a vector that is pass by reference*/
	template<class LD>
	void mimes::util::logspace(LD min, LD max, unsigned int length, std::vector<LD> &X)

    /*logspace where the list returned as a vector*/
	template<class LD>
	std::vector<LD> mimes::util::logspace(LD min, LD max, unsigned int length)
\end{cpp}
Both of these functions work as \cppin{linspace}.

\subsection{{\tt map} function}
The function {\tt map}, defined in {\tt \mimes/src/util/map.hpp}, applys a function of an \cppin{std::vector}, and fills another \cppin{std::vector} with the result. The signatures of this function are

\begin{cpp}
    /*The list is pushed in a vector that is pass by reference*/
	template<class LD>
	void map(const std::vector<LD> &X, std::function<LD(LD)> func,  std::vector<LD> &Y )

    /*The list returned as a vector*/
	template<class LD>
	std::vector<LD> map(const std::vector<LD> &X, std::function<LD(LD)> func)
\end{cpp}
Both of these functions evaluate \cppin{func} (a callable object) for every element of \cppin{X}. If another \cppin{std::vector} is given, then it is cleared and filled with the results, otherwise the function returns a \cppin{std::vector} with the values of  \cppin{func}$($\cppin{X}$)$.

\section{Quick guide to the user input}\label{app:usr_input}
\setcounter{equation}{0}
We present tables~\refs{tab:AxionSolve-input,tab:AxionMass-input,tab:input,tab:template-arguments}, with the various available run-time inputs, required files, and template arguments. In table~\ref{tab:compile_time-options} we show the available compile-time options, that can be used when compiling using the various {\tt makefile} files. 
\begin{table}[h!]
	\centering
	\begin{tabular}{l l}
		\hline\\[-0.4cm]
		\multicolumn{2}{c}{\bf User run-time input for solving the EOM~\ref{eq:eom_sys}.}  \\
		\hline\\[-0.4cm]

		{\tt theta\_i} & Initial angle.  \\
		\hline\\[-0.4cm]

		{\tt fa} & The PQ scale.\\
		\hline\\[-0.4cm]

		{\tt umax } & Once $u=\log a/a_i>${\tt umax}, the integration stops. Typical value: $\sim 500$.\\
		\hline\\[-0.4cm]

		{\tt TSTOP} & Once $T<${\tt TSTOP}, integration stops. Typical value: $10^{-4}~\GeV$.\\
		\hline\\[-0.4cm]

 		{\tt ratio\_ini}& Integration starts at $u$ with $3H/\maT \approx${\tt ratio\_ini}. Typical value: $\sim 10^{3}$.\\
		\hline\\[-0.4cm]

		\multirow{1}{4cm}{{\tt N\_convergence\_max} {\tt convergence\_lim}} & \multirow{1}{12cm}{Integration stops when  the relative difference 
		between two consecutive peaks  is less than {\tt convergence\_lim} for {\tt N\_convergence\_max} 
		consecutive peaks. } \\ \\ \\ 
		\hline\\[-0.4cm]

		{\tt inputFile} & \multirow{1}{12cm}{Relative (or absolute) path to a file that describes the cosmology. The columns should be: $u$ $T ~[\GeV]$ $\log H$, with acceding $u$. Entropy injection should have stopped before the lowest temperature of given in {\tt inputFile}.} \\ \\  \\ \\
		\hline\\[-0.4cm]

		{\tt axionMass} &\multirow{1}{12cm}{Instance of \cppin{mimes::AxionMass<LD>} class. In \CPP this instance is passed as a pointer to the constructor
		of the \cppin{mimes::Axion<LD,Solver,Method>} class, while in \PY it is simply passed as a variable.}\\ \\  \\ \\
		\hline\\[-0.4cm]

		{\tt initial\_stepsize} &  Initial step-size of the solver. Default value: $10^{-2}$.\\ 
		\hline\\[-0.4cm]

		{\tt minimum\_stepsize} & Lower limit of the step-size. Default value:  $10^{-8}$.\\
		\hline\\[-0.4cm]

		{\tt maximum\_stepsize} & Upper limit of the step-size. Default value:  $10^{-2}$.\\
		\hline\\[-0.4cm]

		{\tt absolute\_tolerance} & \multirow{1}{12cm}{Absolute tolerance of the RK solver	(see also table~\ref{tab:RK_mimes_params}).  Default value:  $10^{-8}$.}\\\\
		\hline\\[-0.4cm]

		{\tt relative\_tolerance} & \multirow{1}{12cm}{Relative tolerance of the RK solver	(see also table~\ref{tab:RK_mimes_params}).  Default value:  $10^{-8}$.}\\\\
		\hline\\[-0.4cm]
		
		{\tt beta} & \multirow{1}{12cm}{Aggressiveness of the adaptation strategy	(see also table~\ref{tab:RK_mimes_params}).  Default value:  $0.9$.}\\\\
		\hline\\[-0.4cm]

		{\tt fac\_max}, {\tt fac\_min} &\multirow{1}{12cm}{The step-size does not change more than {\tt fac\_max} and less than {\tt fac\_min} within a trial step (see also table~\ref{tab:RK_mimes_params}). Default values: $1.2$ and $0.8$, respectively.} \\ \\ \\ 
		\hline\\[-0.4cm]
		
		{\tt maximum\_No\_steps} & \multirow{1}{12cm}{If integration needs more than {\tt maximum\_No\_steps} integration stops. Default value: $10^7$.}\\\\
		\hline\\[-0.4cm]
	\end{tabular}
	\caption{Table of the constructor arguments of the \cppin{mimes::AxionSolve<LD,Solver,Method>} class.}
	\label{tab:AxionSolve-input}
\end{table}

\begin{table}[h!]
	\centering
	\begin{tabular}{l l}
		\multicolumn{2}{c}{\bf Input required for the definition of the axion mass via a data file.}  \\
		\hline\\[-0.4cm]
		
		{\tt chi\_Path}& \multirow{1}{12cm}{Relative or absolute path to data file with $T$ (in $\GeV$), $\chi(T)$ (in $\GeV^4$).}\\\\		
		\hline\\[-0.4cm]
		
		{\tt minT} and {\tt maxT}& \multirow{1}{12cm}{Interpolation limits. These are used in order to stop the interpolation at the closest temperatures that exist in the data file. This means that the actual interpolation limits are $T_{\min}\geq${\tt minT} and $T_{\max}\leq${\tt maxT}. Beyond these limits hat axion mass is assumed to be constant. However, this can be changed using 
			\cppin{mimes::AxionMass<LD>::set_ma2_MIN(std::function<LD(LD,LD)> ma2_MIN)}  and \cppin{mimes::AxionMass<LD>::set_ma2_MAX(std::function<LD(LD,LD)> ma2_MAX)}. }\\\\\\\\\\\\\\\\		
		\hline\\[-0.4cm]

		\multicolumn{2}{c}{\bf Input required for the definition of the axion mass via a function.}  \\
		\hline\\[-0.4cm]

		{\tt ma2}& \multirow{1}{12cm}{A function (or a callable object) with signature \cppin{LD ma2(LD T, LD fa)} that returns $\maT^2(T)$.}\\\\		
		\hline\\[-0.4cm]

	\end{tabular}
	\caption{Arguments of the contructors of the \cppin{mimes::AxionMass<LD>} class. }
	\label{tab:AxionMass-input}
\end{table}

\begin{table}[h!]
	\centering
	\begin{tabular}{l l}
		\multicolumn{2}{c}{\bf Required data files, with corresponding variables in {\tt \mimes/Paths.mk}.}  \\
		\hline\\[-0.4cm]
	
		{\tt cosmoDat}& \multirow{1}{12cm}{Relative path to data file with $T$ (in $\GeV$), $\heff$, $\geff$. If the path changes one must run
		{\tt bash configure.sh} and {\tt make}.}\\\\		
		\hline\\[-0.4cm]

		{\tt axMDat}& \multirow{1}{12cm}{Relative path to data file with $T$ (in $\GeV$), $\chi$ (defined from \eqs{eq:axion_mass_def}). If the path changes one must run {\tt bash configure.sh} and {\tt make}. This variable can be omitted, since the user can define an AxionMass instance using any path.}\\\\\\\\		
		\hline\\[-0.4cm]
		
		{\tt anFDat}& \multirow{1}{12cm}{Relative path to data file with $\thetamax$, $f(\thetamax)$. If the path changes one must run
		{\tt bash configure.sh} and {\tt make}.}\\\\		
		\hline\\[-0.4cm]

	\end{tabular}
	\caption{Paths to the required data files. Variables defined in the {\tt \mimes/Paths.mk} files, and used when running {\tt bash configure.sh} in the root directory of \mimes.}
\label{tab:input}
\end{table}

\begin{table}[h!]
	\centering
	\begin{tabular}{l l}
		\multicolumn{2}{c}{\bf Template arguments.}  \\
		\hline\\[-0.4cm]
		
		{\tt LD}& \multirow{1}{12cm}{This template argument appears in all classes of \mimes. The preferred choice is \cppin{long double}. However, in many cases \cppin{double} can be used. The user should be careful, as the later can lead to an inaccurate result; especially for low tolerances, and small values of $\theta$.}\\\\\\\\		
		\hline\\[-0.4cm]
		
		{\tt Solver}& \multirow{1}{12cm}{This is the second template argument of the \cppin{mimes::Axion<LD,Solver,Method>} class. The available choices are {\tt Solver}=$1$ for Rosenbrock method, and {\tt Solver}=$2$ for explicit RK method.}\\\\\\\\
		\hline\\[-0.4cm]
		
		{\tt Method}& \multirow{1}{12cm}{The third template argument of the \cppin{mimes::Axion<LD,Solver,Method>} class. Its value depends on the choice of \cppin{Solver}; For {\tt Solver}=$1$, {\tt Method} can be either \cppin{RODASPR2<LD>} (fourth order) or \cppin{ROS34PW2<LD>} (third order). For {\tt Solver}=$2$, {\tt Method} can only be \cppin{DormandPrince<LD>} (seventh order). Notice that the definitions of the various method classes, also need a template argument,\cppin{LD}, that must be the same as the first template argument of the \cppin{mimes::Axion<LD,Solver,Method>} class. If one defines their own Butcher table (see Appendix~\ref{app:RK}), then they would have to follow their definitions and assumptions.}\\\\\\\\\\\\\\\\\\
		\hline
	\end{tabular}
	\caption{Template arguments of the various \mimes classes.}
	\label{tab:template-arguments}
\end{table}

\begin{table}[h!]
	\centering
	\begin{tabular}{l l}
		\multicolumn{2}{c}{\bf User compile-time options. Variables in the various {\tt Definitions.mk} files.}  \\
		\hline\\[-0.4cm]

		{\tt rootDir}& \multirow{1}{12cm}{The relative path of root directory of \mimes. Relevant only when compiling using {\tt make}. Available in all {\tt Definitions.mk}.}\\\\		
		\hline\\[-0.4cm]
		
		{\tt LONG}& \multirow{1}{12cm}{{\tt long} for \cppin{long double} or empty for \cppin{double}. This is defines a macro in the source files of the various \CPP examples. Available in {\tt Definitions.mk} inside the various subdirectories of {\tt \mimes/UserSpace/Cpp}.}\\\\\\\\		
		\hline\\[-0.4cm]

		{\tt LONGpy}& \multirow{1}{12cm}{{\tt long} or empty. Same as {\tt LONG}, applies in the \PY modules. Available in {\tt \mimes/Definitions.mk}.}\\\\		
		\hline\\[-0.4cm]

		{\tt SOLVER}& \multirow{1}{12cm}{In order to use a Rosenbrock method {\tt SOLVER}=$1$. For explicit RK method, {\tt SOLVER}=$2$. This defines a macro that is passes as the second template argument of \cppin{mimes::Axion<LD,Solver,Method>}.  The corresponding variable in {\tt \mimes/Definitions.mk} applies to the \PY modules. The variable in {\tt \mimes/UserSpace/Cpp/Axion/Definitions.mk} applies to the example in the same directory.}\\\\\\\\\\\\\\		
		\hline\\[-0.4cm]

		{\tt METHOD}& \multirow{1}{12cm}{Depending on the solver, this variable should name one of its available methods. For {\tt SOLVER}=$1$, {\tt METHOD}={\tt RODASPR2}(fourth order) or {\tt ROS34PW2}(third order). For {\tt SOLVER}=$2$, {\tt METHOD}={\tt DormandPrince }(seventh order). There is a macro ({\tt METHOD}) used by the shared library {\tt \mimes/lib/Axion\_py.so}. The corresponding variable in {\tt \mimes/Definitions.mk} applies to the \PY modules. The variable in {\tt \mimes/UserSpace/Cpp/Axion/Definitions.mk} applies to the example in that directory.}\\\\\\\\\\\\\\\\
		 		
		\hline\\[-0.4cm]
		
		\multicolumn{2}{c}{\bf Compiler options}  \\
		\hline\\[-0.4cm]
		
		{\tt CC} &  \multirow{1}{12cm}{The preferred \CPP compiler ({\tt g++} by default). Corresponding variable in all {\tt Definitions.mk} files.} \\\\
		\hline\\[-0.4cm]
		
		{\tt OPT} &  \multirow{1}{12cm}{Available options are {\tt OPT}={\tt O1}, {\tt O2}, {\tt O3} (be default). This variable defines the optimization level of the compiler. The variable can be changed in all {\tt Definitions.mk} files. In the root directory of \mimes, the optimization level applies to the python modules (\ie the shared libraries), while in the subdirectories of {\tt \mimes/UserSpace/Cpp} it only applies to example inside them.}   \\\\\\\\\\\\
		\hline\\[-0.4cm]

	\end{tabular}
	\caption{User compile-time input and options. These are available in the various {\tt Definitions.mk} files, which are used when compiling using {\tt make}.}
	\label{tab:compile_time-options}
\end{table}

\pagebreak
\bibliography{refs}{}
\bibliographystyle{JHEP}                        

\end{document}